\pdfoutput=1
\documentclass[11pt,letterpaper]{article}
\usepackage[margin=1in]{geometry}
\usepackage{amsmath,amsfonts}
\usepackage{algorithmic}
\usepackage{algorithm}
\usepackage{array}
\usepackage{textcomp}
\usepackage{booktabs}
\usepackage{url}
\usepackage{multirow}
\usepackage{verbatim}
\usepackage{graphicx}
\usepackage{cite}
\usepackage{amsmath,amssymb,mathtools}
\usepackage{bm}
\usepackage[hidelinks]{hyperref}
\date{}
 
\begin{document}
 
\title{GeoDiff-SAR II: 3D Model-Guided SAR Image Generation with Explicit Control of Key Imaging Parameters}
 
\author{Xuanting Wu, Fan Zhang, Fei Ma, Yingbing Liu, Lingxiao Peng, Qiang Yin, and Yongsheng Zhou%
\thanks{Xuanting Wu, Fan Zhang, Fei Ma, Yingbing Liu, Qiang Yin, and Yongsheng Zhou are with the College of Information Science and Technology, Beijing University of Chemical Technology, Beijing 100029, China.}%
\thanks{Lingxiao Peng is with the Suzhou Aerospace Information Research Institute, Chinese Academy of Sciences, Suzhou 215123, China.}%
\thanks{Corresponding author: Fei Ma (e-mail: mafei@mail.buct.edu.cn).}} 
 
\maketitle
 
\begin{abstract}
Existing Synthetic Aperture Radar (SAR) image generation methods still lack reliable controllability over key imaging parameters, particularly azimuth angle, depression angle, and polarization mode. Our preliminary GeoDiff-SAR supported limited azimuth completion, but remained ineffective for large missing azimuth sectors and did not provide unified control over multiple imaging conditions. To address this problem, we propose GeoDiff-SAR II, a 3D model-guided decoupled framework for controllable SAR image generation. The proposed framework imposes controllability through physically grounded geometric-electromagnetic cues rather than image intensity alone. We introduce a Geometric-Electromagnetic Conditioning Map (GECM), a structured intermediate representation that encodes the target pose map and dominant scattering centers, thereby decoupling macroscopic geometry from microscopic scattering responses. During training, GECMs are derived from real sparse-azimuth SAR images. During inference, the same representation is rendered directly from a 3D CAD model under specified azimuth, depression angle, and polarization conditions, enabling physically consistent control across large viewpoint gaps. The imaging parameters are further converted into text conditions, while the GECM is injected through ControlNet to provide explicit spatial guidance. Combined with Low-Rank Adaptation (LoRA) on a FLUX backbone, the proposed framework unifies geometric-electromagnetic conditioning and parameter-aware generation within a single process. Experiments on simulated and real datasets demonstrate controllable generation over key SAR imaging parameters, stable generalization across large azimuth gaps, and consistent improvements in image fidelity, physical consistency, and downstream Automatic Target Recognition (ATR) performance.
\end{abstract}
 
\noindent\textbf{Keywords:}
Synthetic Aperture Radar (SAR), Image Generation, Diffusion Models, Physics-Guided, Zero-Shot Learning.
 
\section{Introduction}
\label{sec:introduction}
 
Synthetic Aperture Radar (SAR) provides all-weather, day-and-night imaging and plays an important role in earth observation and Automatic Target Recognition (ATR) \cite{moreira2013tutorial, zhu2021deep}. Recent deep models have substantially improved SAR ATR \cite{lecun2015deep, chen2016target}, but their performance depends on large multi-view annotated datasets. In practice, flight cost, revisit constraints, and confidentiality make full-aspect SAR acquisition difficult \cite{ding2020few, zhang2021sar}. As a result, ATR models often degrade on unseen viewpoints, creating the sparse-azimuth bottleneck \cite{wang2022sparse, liu2023sparse}. 
 
\begin{figure}[!t]
\centering
\includegraphics[width=0.6\linewidth]{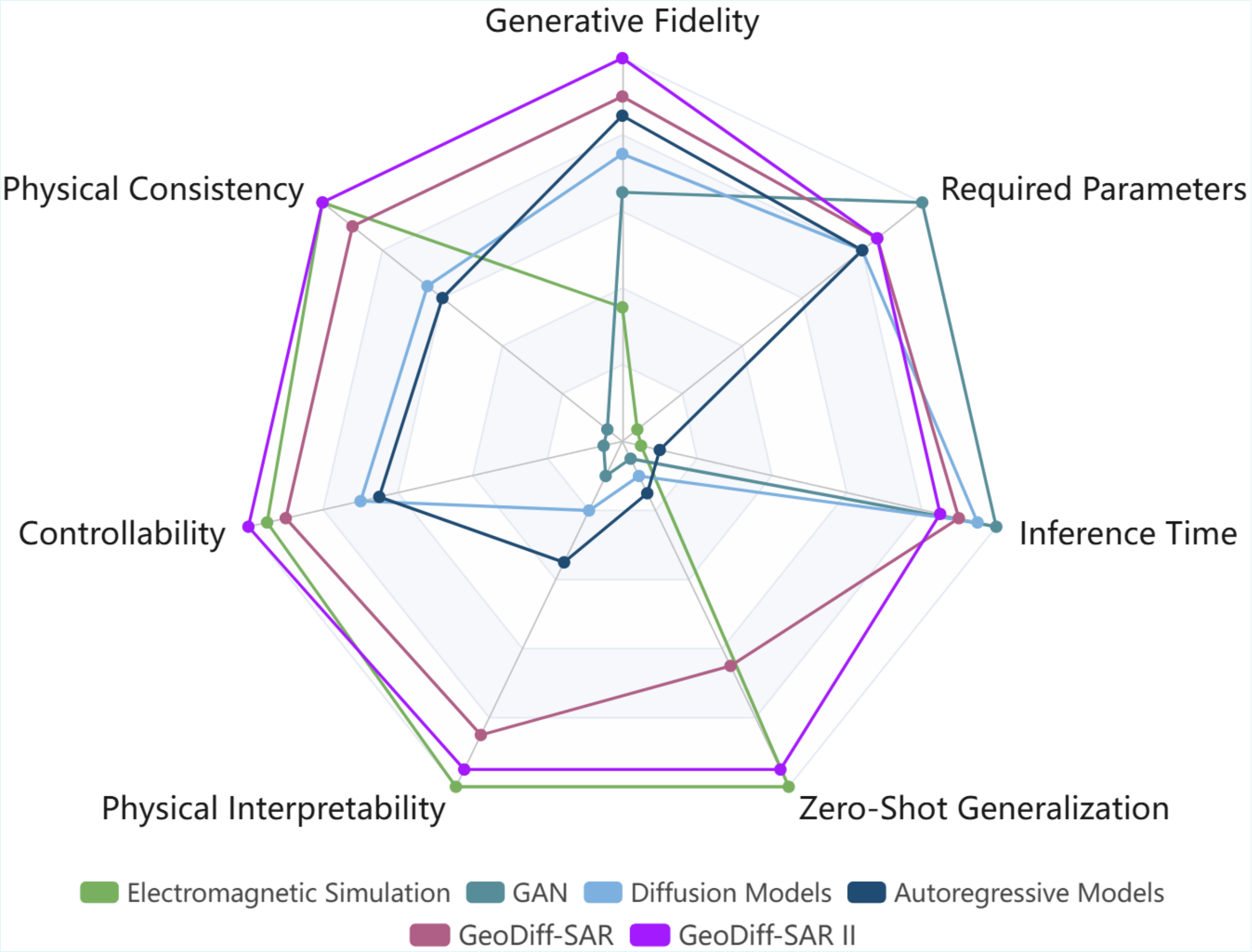}
\caption{Comprehensive comparison of representative generation paradigms across multiple capabilities. Traditional electromagnetic simulations offer high physical consistency but limited generative fidelity, whereas purely data-driven deep learning models (e.g., GANs and diffusion models) can achieve high realism but offer limited physical interpretability under zero-shot conditions. Compared with our preliminary GeoDiff-SAR, GeoDiff-SAR II is intended to better balance physical fidelity, zero-shot generalization, and generative realism.}
\label{fig:intro_radar}
\end{figure}

Generative modeling offers a promising means of alleviating SAR data scarcity. However, although GANs \cite{goodfellow2014generative}, large conditional generators \cite{ramesh2022hierarchical}, and diffusion models \cite{ho2020denoising, rombach2022high} have achieved strong performance in optical synthesis, transferring them directly to SAR remains difficult \cite{guo2021sar, ji2023diffusion}. SAR imagery is governed by microwave scattering and exhibits foreshortening, layover, shadowing, and multi-bounce effects \cite{cumming2005digital}. As a result, existing SAR generation methods still struggle to provide controllable synthesis with respect to key imaging parameters such as azimuth angle, depression angle, and polarization mode. Without explicit physical constraints, generated samples may exhibit physically inconsistent artifacts, referred to in recent work as ``physical hallucinations'' \cite{zhao2023physical}, which can further affect downstream ATR \cite{xia2024hallucination, wang2023hallucination}. Figure~\ref{fig:intro_radar} summarizes these trade-offs and motivates GeoDiff-SAR II.

Our preliminary GeoDiff-SAR \cite{zhang2026geodiff} demonstrated that geometric priors can improve sparse-view SAR generation and partially support azimuth completion. Nevertheless, its controllability remained limited to a relatively narrow range and was insufficient when large azimuth sectors were absent from the training data. Furthermore, it did not establish a unified mechanism for jointly controlling azimuth angle, depression angle, and polarization mode within a physically consistent generation process.

Inferring unseen 3D configurations directly from 2D SAR intensities is fundamentally ill-posed, especially under large viewpoint changes. This motivates the introduction of a 3D CAD model as a physically meaningful prior \cite{kim20203d}. To bridge macroscopic structure and microscopic scattering behavior, we introduce a Geometric-Electromagnetic Conditioning Map (GECM), which integrates the target pose skeleton and dominant scattering centers into a deterministic intermediate representation. Notably, this representation can be derived from real SAR images during training or rendered directly from a 3D model during inference, thereby providing a unified geometric-electromagnetic bridge between data-driven supervision and physics-driven controllability.

\begin{figure}[!t]
\centering
\includegraphics[width=0.6\linewidth]{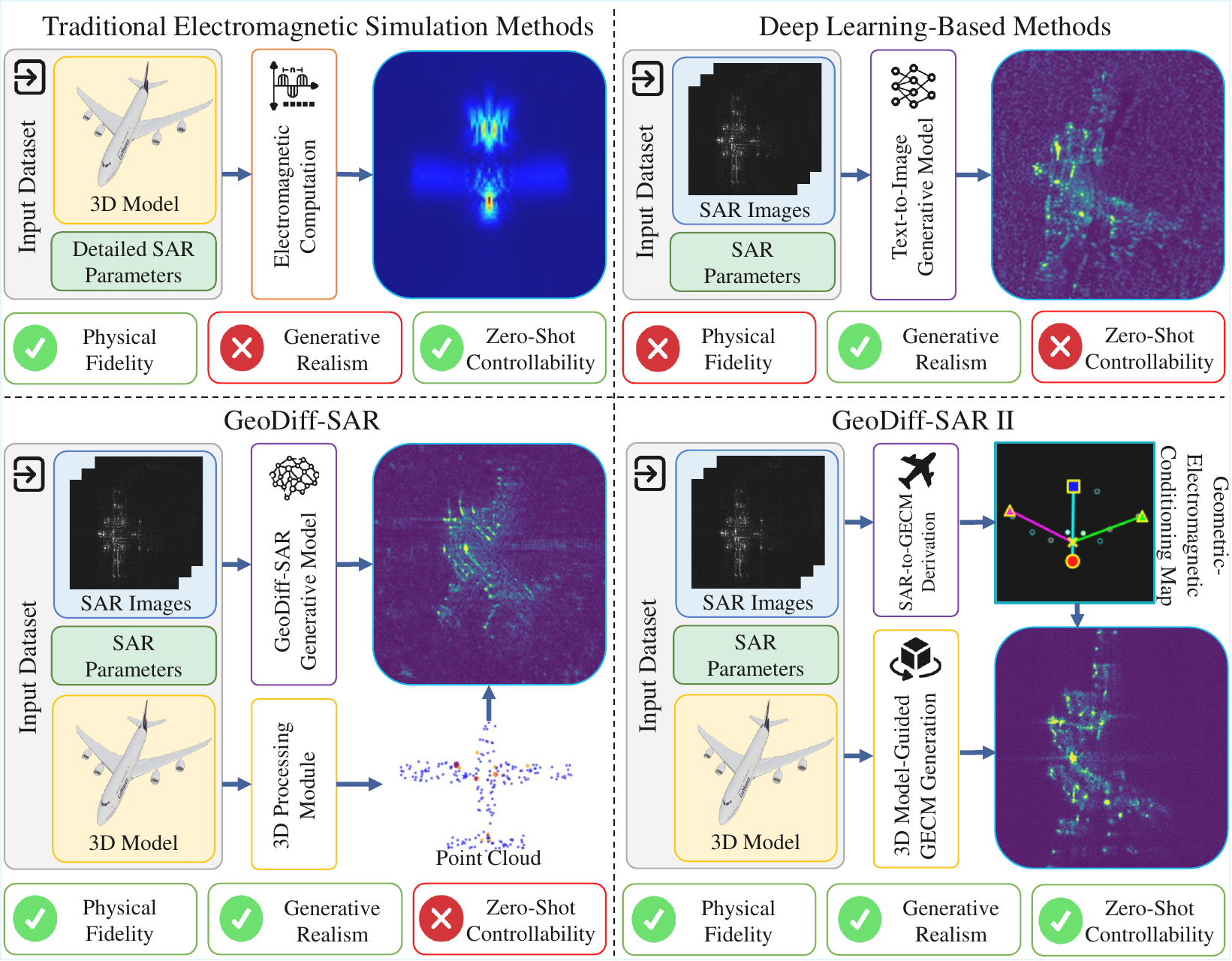}
\caption{Evolution of SAR image generation frameworks. (a) Traditional electromagnetic simulations follow physical principles but suffer from a substantial domain gap. (b) Purely data-driven models learn realistic distributions but offer limited physical interpretability and may exhibit geometric inconsistency. (c) Our predecessor, GeoDiff-SAR, introduced 3D point cloud priors but relied on implicit feature fusion. (d) GeoDiff-SAR II explicitly decouples generation into a deterministic physical forward engine (yielding the GECM) and a stochastic texture rendering network to improve realism, zero-shot controllability, and physical consistency.}
\label{fig:intro_cmp}
\end{figure}

Based on GECM, we build a 3D model-guided decoupled SAR generation framework. During training, a SAR-to-GECM derivation pipeline produces ground-truth GECMs from real sparse-azimuth SAR data. During inference, a physics-informed 3D forward engine renders synthetic GECMs from CAD models under specified azimuth, depression angle, and polarization modes. The resulting GECM is injected through ControlNet \cite{zhang2023adding} to impose explicit spatial control, while SAR metadata are converted into text conditions and optimized through Low-Rank Adaptation (LoRA) modules within a FLUX backbone \cite{blackforest2024flux}. Compared with our preliminary GeoDiff-SAR \cite{zhang2026geodiff}, which relied on implicit fusion, this explicit decoupling of geometric-electromagnetic conditioning and text-guided appearance generation is intended to improve texture-structure alignment and zero-shot generalization, as illustrated in Fig.~\ref{fig:intro_cmp}.

The main contributions of this article are summarized as follows:
\begin{itemize}
    \item We introduce a geometric-electromagnetic feature derivation strategy centered on GECM, which extracts pose structure and dominant scattering centers either from real SAR images or directly from 3D CAD models, thereby unifying image-driven supervision and 3D-model-driven controllability.
    \item We propose a decoupled SAR generation framework that jointly incorporates GECM-based geometric-electromagnetic conditioning and metadata-derived text control into a ControlNet- and LoRA-enhanced foundation model, enabling explicit control over key SAR imaging parameters.
    \item We conduct extensive experiments on both simulated and real datasets, demonstrating that the proposed method achieves controllable generation across azimuth angle, depression angle, and polarization mode, maintains performance under large azimuth gaps, and consistently improves both image quality and downstream ATR performance.
\end{itemize}
 
\section{Related Work}
\label{sec:related_work}
 
\subsection{SAR Image Generation and Data Augmentation}
The difficulty of collecting multi-view SAR data has long motivated research on data augmentation and image synthesis \cite{moreira2013tutorial, zhu2021deep}. Traditional approaches rely on electromagnetic simulation, such as SBR and PO \cite{franceschetti2003sar, auer2009ray}. Although physically interpretable, these methods remain computationally expensive and often suffer from a sim-to-real gap in clutter, speckle, and sensor artifacts \cite{schreiber2005scattering, lewis2019generative}. 
 
With the rise of deep learning, data-driven generators have become an alternative to physics-based synthesis. GAN-based frameworks, including Pix2Pix \cite{isola2017image} and CycleGAN \cite{zhu2017unpaired}, have been used to translate semantic or optical priors into SAR \cite{guo2021sar}. However, GANs remain sensitive to optimization instability and provide only limited geometric control under sparse-view conditions \cite{arjovsky2017wasserstein}.
 
Beyond adversarial learning, discrete-token and large-scale conditional generators \cite{esser2021taming, ramesh2022hierarchical} further expanded the design space, but they do not resolve the viewpoint-dependent physics of SAR formation.
 
Recently, diffusion models \cite{ho2020denoising, rombach2022high} have become a widely adopted framework for high-fidelity image synthesis. Early SAR studies explored diffusion-based generation \cite{ji2023diffusion}, yet generic diffusion models still treat SAR generation mainly as 2D appearance modeling and remain vulnerable to geometric inconsistency under sparse-azimuth and zero-shot conditions \cite{xia2024hallucination, zhao2023physical, wang2023hallucination}.
 
\subsection{Physics-Informed Deep Learning in SAR}
To mitigate ambiguity in 2D SAR appearance, recent work has introduced physical priors into deep models \cite{karniadakis2021physics}. Early SAR studies mainly used 2D cues such as masks or edges \cite{malmgren2017improving}. Although useful, these planar constraints do not adequately capture viewpoint-dependent topological changes caused by Line-of-Sight (LOS), layover, or multipath scattering \cite{zhong2025scattering}.
 
Our preliminary GeoDiff-SAR (V1) \cite{zhang2026geodiff} incorporated simulated 3D point clouds as geometric priors and fused them with text and image features through Feature-wise Linear Modulation (FiLM) \cite{perez2018film}. However, implicit modulation remained insufficient for reliable spatial control under highly unseen viewpoints, and the physical branch was limited to single-bounce contours. These limitations motivate a framework with explicit spatial conditioning and richer scattering-aware priors.
 
\subsection{Controllable Diffusion Models}
Recent controllable diffusion and foundation models offer practical mechanisms for structured generation. LoRA \cite{hu2021lora} enables SAR-domain adaptation of large backbones such as DiT \cite{peebles2023scalable} and FLUX \cite{blackforest2024flux} with moderate cost, while ControlNet \cite{zhang2023adding} provides explicit spatial controllability.
 
The key challenge in SAR is that paired multi-view physical condition maps are rarely available from real data. GeoDiff-SAR II addresses this gap by combining a SAR-to-GECM derivation process for real training images with a 3D model-guided forward engine for test-time rendering.
 
\begin{figure*}[!t]
\centering
\includegraphics[width=\textwidth]{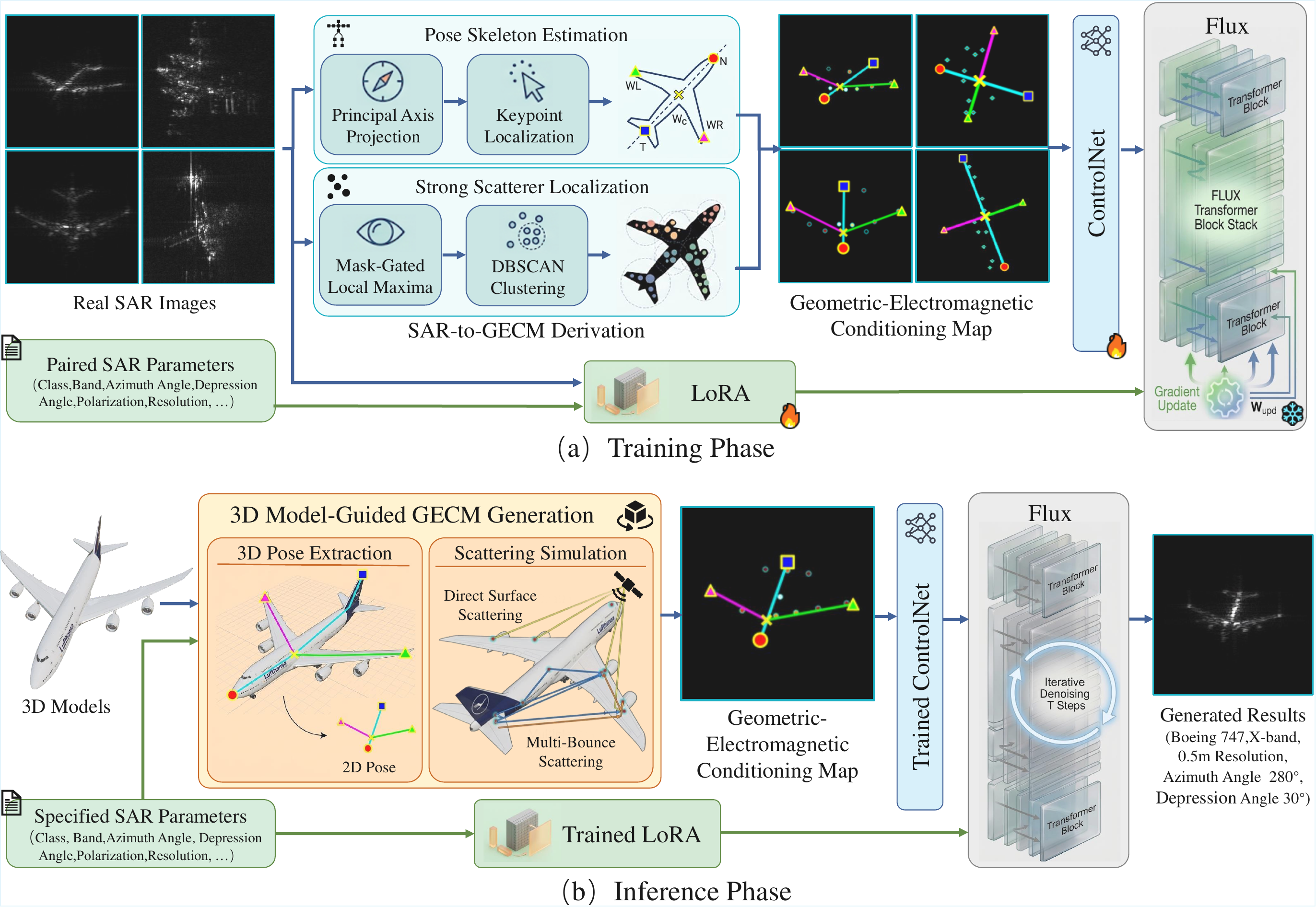}
\caption{Overall architecture of the proposed 3D Model-Guided decoupled SAR image generation framework. (a) \textbf{Training Phase}: A SAR-to-GECM derivation process infers spatial conditions from real sparse-azimuth SAR images. It utilizes a \emph{Pose Skeleton Estimation} branch (via principal axis projection) and a \emph{Strong Scatterer Localization} branch (via mask-gated local maxima and DBSCAN) to render the ground-truth GECM. This GECM serves as an explicit spatial constraint for ControlNet, while paired SAR parameters optimize LoRA modules injected into the FLUX 1.0 backbone to learn sensor-specific textures. (b) \textbf{Inference Phase}: Operating in a zero-shot manner, a 3D Model-Guided GECM Generation engine performs forward physical simulation based on a 3D CAD model and specified imaging parameters. Through 3D pose extraction and multi-bounce scattering simulation, it renders a synthetic GECM via a continuous-discrete asymmetric projection. This simulated GECM and metadata-derived text conditions subsequently drive the trained joint network to synthesize SAR images without relying on real target masks.}
\label{fig:overall_architecture}
\end{figure*}
 
\section{Method}
\subsection{Overall Architecture}
\label{subsec:overall_architecture}
 
To address physical hallucinations in zero-shot sparse-azimuth SAR generation, we formulate synthesis as two coupled but explicit sub-tasks: macroscopic structural conditioning and microscopic electromagnetic texture rendering. 

As illustrated in Fig.~\ref{fig:overall_architecture}, the framework consists of a real-data training phase and a 3D-guided zero-shot inference phase, both connected by the GECM.

\paragraph{Training Phase: SAR-to-GECM Derivation and Joint Optimization}
During training (Fig.~\ref{fig:overall_architecture}(a)), the network learns a mapping from GECM-based spatial conditions to realistic SAR appearance. A dedicated SAR-to-GECM derivation process recovers the pose skeleton and dominant scattering centers from real sparse-azimuth SAR images, and the resulting control map is injected through ControlNet \cite{zhang2023adding}. In parallel, SAR metadata optimizes LoRA \cite{hu2021lora} modules so that the backbone learns sensor-specific textures and clutter statistics.
 
\paragraph{Inference Phase: 3D Model-Guided Zero-Shot Generation}
During inference (Fig.~\ref{fig:overall_architecture}(b)), real SAR images are no longer required. Instead, a CAD model and imaging parameters drive a physics-informed forward engine (Section~\ref{sec:3d_gecm_generation}) to render a synthetic GECM consistent with the training semantics. The simulated GECM and metadata-derived text condition then guide iterative denoising to synthesize SAR images consistent with both the target geometry and the underlying scattering mechanism.
 
\subsection{SAR-to-GECM Derivation}
\label{subsec:gecm_extractor}
 
Given a real SAR image $I \in \mathbb{R}^{H\times W}$, we construct a GECM mathematically defined as
\begin{equation}
\mathcal{G} = \left(\mathbf{P}_{\text{pose}}, \mathbf{Q}_{\text{scat}}\right),
\end{equation}
where $\mathbf{P}_{\text{pose}}=\{\mathbf{p}_n, \mathbf{p}_t, \mathbf{p}_w, \mathbf{p}_l, \mathbf{p}_r\}$ denotes the spatial coordinates of the aircraft's nose, tail, wing-root, left-wing tip, and right-wing tip, respectively. The term $\mathbf{Q}_{\text{scat}}=\{(\mathbf{q}_k, \alpha_k)\}_{k=1}^{K}$ denotes a set of $K$ strong scattering centers with their corresponding normalized intensities $\alpha_k$.
 
\begin{figure*}[!t]
\centering
\includegraphics[width=\textwidth]{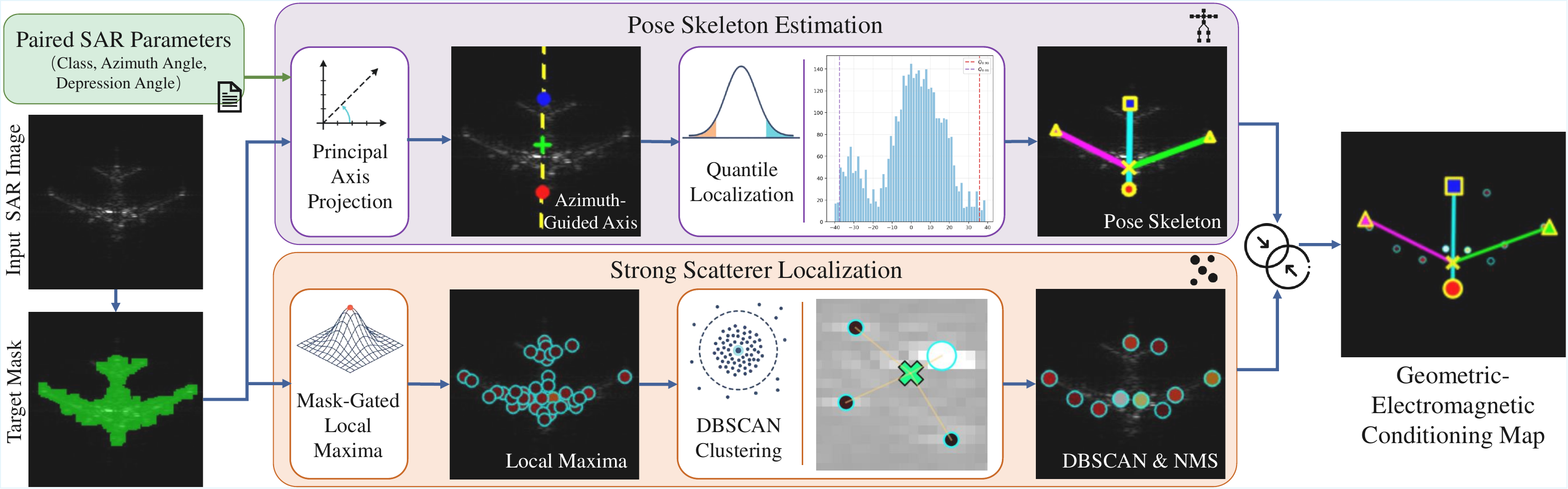}
\caption{Detailed pipeline of SAR-to-GECM derivation. From a real SAR image and its corresponding parameters, the target mask is first extracted. The Pose Skeleton Estimation branch determines the macroscopic skeleton via principal axis projection and quantile localization. Concurrently, the Strong Scatterer Localization branch isolates microscopic electromagnetic responses via mask-gated local maxima detection and DBSCAN clustering, ultimately yielding the ground-truth GECM.}
\label{fig:gecm_extractor}
\end{figure*}
 
As illustrated in Fig.~\ref{fig:gecm_extractor}, intensity normalization is first applied to standardize the input. For high-dynamic-range inputs, a log compression $I_{\log} = \log(1+I)$ is utilized. The image is then converted to grayscale and normalized to $[0,1]$:
\begin{equation}
\tilde{I} = \frac{I - \min(I)}{\max(I)-\min(I)+\epsilon_0}.
\end{equation}
To isolate the target from background clutter, a binary mask $M \in \{0,1\}^{H\times W}$ is estimated through CLAHE enhancement, Otsu thresholding, and morphological filtering. We select the optimal connected component by evaluating a scoring function that penalizes the spatial distance from the image center $\mathbf{u}_0$. Specifically, $M = \mathcal{D}(\mathcal{M}_{c^*})$, where $\mathcal{D}(\cdot)$ denotes morphological dilation, and $\mathcal{M}_{c^*}$ is the binary mask of the component $c^*$ satisfying:
\begin{equation}
c^* = \arg\max_{c\in \mathcal{K}} \left[\operatorname{Area}(c)-\lambda\|\mathbf{u}_c-\mathbf{u}_0\|_2\right],
\end{equation}
where $\mathcal{K}$ represents the set of all connected components, and $\mathbf{u}_c$ is the centroid of component $c$.
 
Based on the derived mask $M$, we determine the pose skeleton guided by an azimuth prior. Let the foreground pixels be $\mathcal{P}=\{(x_i,y_i)\mid M(x_i,y_i)=1\}$. The spatial centroid $\mathbf{c}$ is calculated as:
\begin{equation}
\mathbf{c} = \frac{1}{|\mathcal{P}|}\sum_{(x_i,y_i)\in\mathcal{P}}(x_i,y_i).
\end{equation}
Given the azimuth angle $\alpha$ (following the dataset convention where $0^\circ$ points West with clockwise increase), the unit directional vector for the nose in image coordinates is defined as:
\begin{equation}
\mathbf{d}_n = (-\cos\alpha,\,-\sin\alpha).
\end{equation}
If the azimuth is unavailable, a Principal Component Analysis (PCA) major-axis fallback is employed. The orthogonal lateral direction is determined as $\mathbf{d}_l = (d_{n,y},\, -d_{n,x})$. By projecting each foreground point $\mathbf{p}_i\in\mathcal{P}$ onto these axes, we obtain the longitudinal and lateral projections:
\begin{equation}
a_i = (\mathbf{p}_i-\mathbf{c})^\top \mathbf{d}_n, \quad s_i = (\mathbf{p}_i-\mathbf{c})^\top \mathbf{d}_l.
\end{equation}
The front and back extents are robustly estimated via quantiles $\ell_f = Q_{0.99}(\{a_i\})$ and $\ell_b = -Q_{0.01}(\{a_i\})$, yielding the longitudinal keypoints:
\begin{equation}
\mathbf{p}_n=\mathbf{c}+\ell_f\mathbf{d}_n,\qquad \mathbf{p}_t=\mathbf{c}-\ell_b\mathbf{d}_n.
\end{equation}
Setting the wing root at the centroid ($\mathbf{p}_w=\mathbf{c}$), we isolate a longitudinal band $a_i\in(-0.2\ell_b,\,0.3\ell_f)$ to estimate the left and right wing lengths via the $98$-th side quantiles ($Q_{0.98}$). These are clipped to the image bounds, yielding the left and right wing tips $\mathbf{p}_l$ and $\mathbf{p}_r$, and thus defining the final pose skeleton $\mathbf{P}_{\text{pose}}=\{\mathbf{p}_n, \mathbf{p}_t, \mathbf{p}_w, \mathbf{p}_l, \mathbf{p}_r\}$.
 
Parallel to pose-skeleton estimation, we localize candidate strong scatterers. The normalized intensity $\tilde{I}$ is smoothed via a Gaussian filter $I_s = \mathcal{G}_\sigma * \tilde{I}$. A mask-aware adaptive threshold is defined as $\tau = Q_{0.90}(I_s \mid M=1)$. Candidate peaks are identified using a local-maxima constraint combined with mask gating:
\begin{equation}
\mathcal{C}=\left\{(x,y)\,\middle|\, 
\begin{aligned}
& I_s(x,y)\ge \mathcal{D}(I_s)(x,y)-\delta,\; \\
& I_s(x,y)\ge\tau,\; M(x,y)=1 
\end{aligned}
\right\}.
\end{equation}
If no candidates are found ($\mathcal{C}=\emptyset$), a global fallback threshold $Q_{0.97}(I_s)$ is triggered. The candidates are ranked by intensity and filtered using distance-constrained Non-Maximum Suppression (NMS).
 
To refine these candidates into representative scattering centers, let the candidate set with associated intensities be denoted as $\mathcal{C}=\{(x_j,y_j, s_j)\}_{j=1}^{N_c}$. We apply the DBSCAN clustering algorithm on the spatial coordinates $\mathbf{z}_j=(x_j,y_j)$ to yield cluster labels $\{l_j\}=\operatorname{DBSCAN}(\{\mathbf{z}_j\};\varepsilon,m)$. For each valid cluster $k$, the intensity-weighted center $(\bar{x}_k, \bar{y}_k)$ is computed as:
\begin{equation}
\bar{x}_k=\frac{\sum_{j:l_j=k} w_j x_j}{\sum_{j:l_j=k} w_j},\qquad \bar{y}_k=\frac{\sum_{j:l_j=k} w_j y_j}{\sum_{j:l_j=k} w_j},
\end{equation}
where $w_j=s_j+\epsilon_1$, and the cluster's representative intensity is $\bar{s}_k=\max_{j:l_j=k} s_j$. If DBSCAN yields no valid clusters, the initial NMS-selected candidates are utilized as a fallback. After retaining a small number of high-score noise points, a final NMS enforces spatial diversity and caps the output to $K$ salient points, forming the final scatterer set $\mathbf{Q}_{\text{scat}}=\{(\mathbf{q}_k, \alpha_k)\}_{k=1}^{K}$, where $\mathbf{q}_k$ denotes the spatial coordinates and $\alpha_k$ the corresponding intensity.
 
\begin{figure}[!t]
\centering
\includegraphics[width=0.4\linewidth]{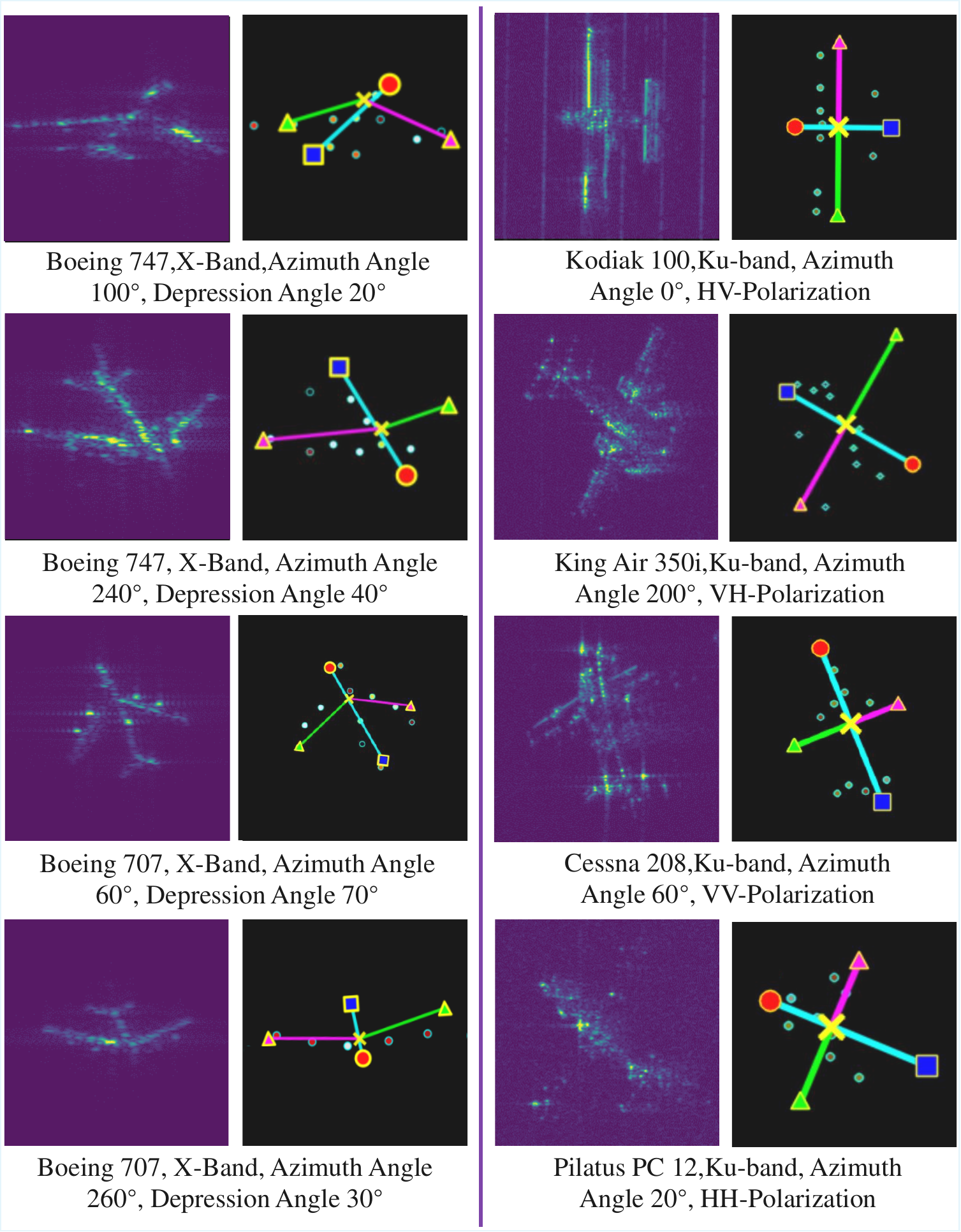}
\caption{Visualized examples of ground-truth GECMs derived from real SAR images across various aircraft classes, frequency bands, and polarization modes.}
\label{fig:gecm_extractor_result}
\end{figure}

Finally, the derived structural and electromagnetic cues are aggregated to render the GECM. The map is drawn on a black canvas incorporating the body axis, nose/tail, wing markers, the wing-root connector, and the strong scatter points $\mathbf{Q}_{\text{scat}}$. The color semantics are fixed to match the training control style, ensuring consistency between the training and inference phases. Representative GECMs derived from real SAR images are shown in Fig.~\ref{fig:gecm_extractor_result}.

\subsection{3D Model-Guided GECM Generation}
\label{sec:3d_gecm_generation}
 
\begin{figure*}[!t]
\centering
\includegraphics[width=\textwidth]{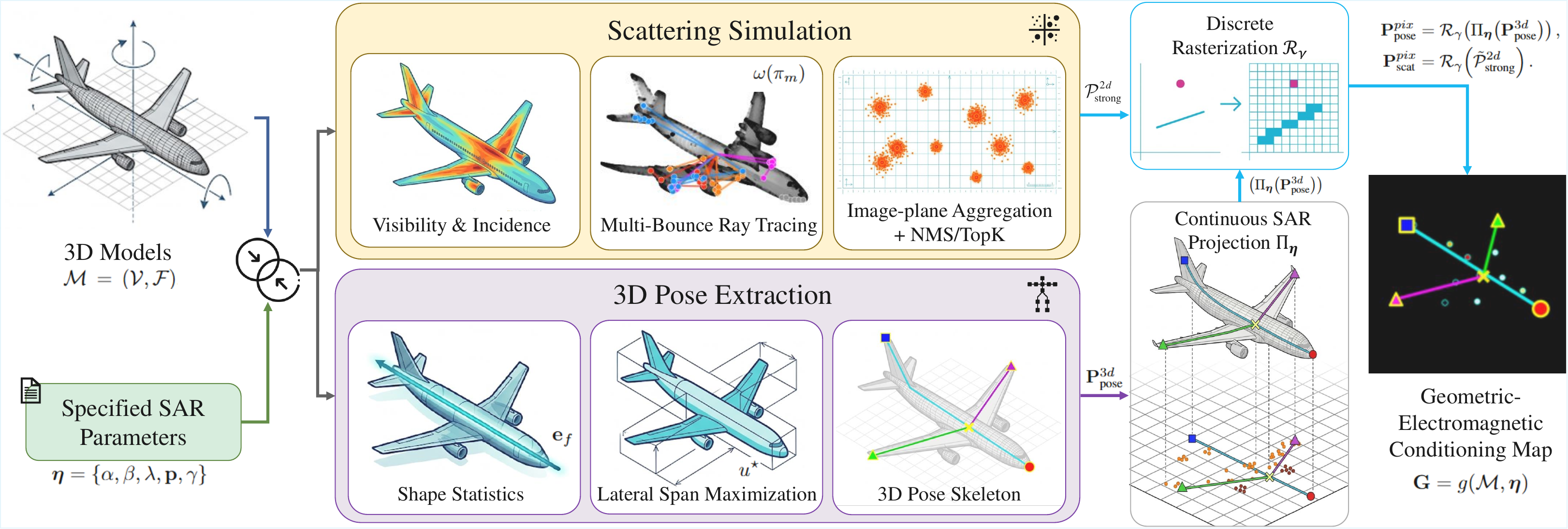}
\caption{Detailed pipeline of the 3D Model-Guided GECM Generation. A 3D CAD mesh is transformed into the radar coordinate system based on specified imaging parameters. The 3D Pose Extraction module localizes the structural skeleton, while the Scattering Simulation module performs multi-bounce ray tracing. The resulting strong scatterers undergo spatial aggregation within a continuous image plane before merging with the projected 3D skeleton through a discrete rasterization operator ($\mathcal{R}_{\gamma}$), rendering the final zero-shot GECM.}
\label{fig:3d_gecm_pipeline}
\end{figure*}

We propose a 3D model-guided GECM generation method, a physics-informed conditioning procedure that converts a known aircraft CAD mesh and specified imaging parameters into a structured GECM for zero-shot ControlNet-guided inference. As illustrated in Fig.~\ref{fig:3d_gecm_pipeline}, the key motivation is that SAR image formation is a many-to-one 2D projection that depends strongly on viewpoint; therefore, directly inferring stable geometric semantics from SAR intensity alone is ill-posed. By explicitly injecting a 3D model, we constrain the pose topology and scattering mechanisms in a physically consistent manner, thereby reducing the ambiguity of downstream generation.
 
Let the aircraft CAD mesh be denoted as $\mathcal{M}=(\mathcal{V},\mathcal{F})$, comprising vertices $\mathbf{v}\in\mathbb{R}^3$ and triangular facets $f\in\mathcal{F}$. The specified SAR imaging parameters are defined as:
\begin{equation}
\label{eq:params}
\boldsymbol{\eta}=\{\alpha, \beta, \lambda, \mathbf{p}, \gamma\},
\end{equation}
where $\alpha$ is the azimuth angle, $\beta$ is the depression angle, $\lambda$ is the radar wavelength, $\mathbf{p}\in\{\mathrm{HH},\mathrm{HV},\mathrm{VH},\mathrm{VV}\}$ is the polarization channel, and $\gamma$ is the spatial resolution. We first transform the mesh into the radar coordinate frame:
\begin{equation}
\label{eq:transform}
\mathbf{v}' = \mathbf{R}_{\mathrm{az}}(\alpha)\,\mathbf{R}_{\mathrm{dep}}(\beta)\,\mathbf{v}.
\end{equation}
In implementation, global translation is absorbed by subsequent projection-to-canvas normalization; hence it is omitted in Eq.~(\ref{eq:transform}) for brevity. Following the dataset convention where azimuth $0^\circ$ points to geographic West and increases clockwise, the azimuth rotation matrix is formulated as:
\begin{equation}
\label{eq:rotation}
\mathbf{R}_{\mathrm{az}}(\alpha)=
\begin{bmatrix}
\cos\alpha & \sin\alpha & 0\\
-\sin\alpha & \cos\alpha & 0\\
0 & 0 & 1
\end{bmatrix}.
\end{equation}
Consequently, the incident unit look vector is derived as:
\begin{equation}
\label{eq:look_vector}
\hat{\mathbf{k}}_{\mathrm{i}}=
\begin{bmatrix}
\sin\alpha\cos\beta\\
\cos\alpha\cos\beta\\
-\sin\beta
\end{bmatrix}.
\end{equation}
 
To extract the macroscopic geometry, the aircraft skeleton is computed from the transformed 3D mesh $\mathcal{V}'$ using a geometry-driven pipeline. We first estimate a fuselage-forward axis $\mathbf{e}_f$ via PCA, capturing the principal shape direction, and then build a 3D centerline $\mathcal{C}_f(u)$ by calculating slice medians along $\mathbf{e}_f$. The lateral axis $\mathbf{e}_s$ defines the left/right wing directions. The wing root (joint) is initialized at the slice with maximal lateral support and is subsequently constrained to lie on the fuselage centerline:
\begin{equation}
\label{eq:wing_root}
\begin{split}
u^\star &= \mathop{\arg\max}_{u} \big(\mathrm{span}_L(u)+\mathrm{span}_R(u)\big),\\
\mathbf{x}_{\mathrm{joint}} &= \Pi_{\mathcal{C}_f}\!\big(\mathcal{C}_f(u^\star)\big),
\end{split}
\end{equation}
where $\Pi_{\mathcal{C}_f}(\cdot)$ denotes the projection operator onto the fuselage centerline. The wing tips are selected from extreme support points along $\pm\mathbf{e}_s$ and are regularized by contour-consistency and bilateral-length balancing. The nose and tail are defined as the two endpoints of $\mathcal{C}_f(u)$ after semantic orientation correction. These five keypoints form the 3D pose skeleton set $\mathbf{P}_{\text{pose}}^{3d}$.
 
\begin{figure}[!t]
\centering
\includegraphics[width=0.5\linewidth]{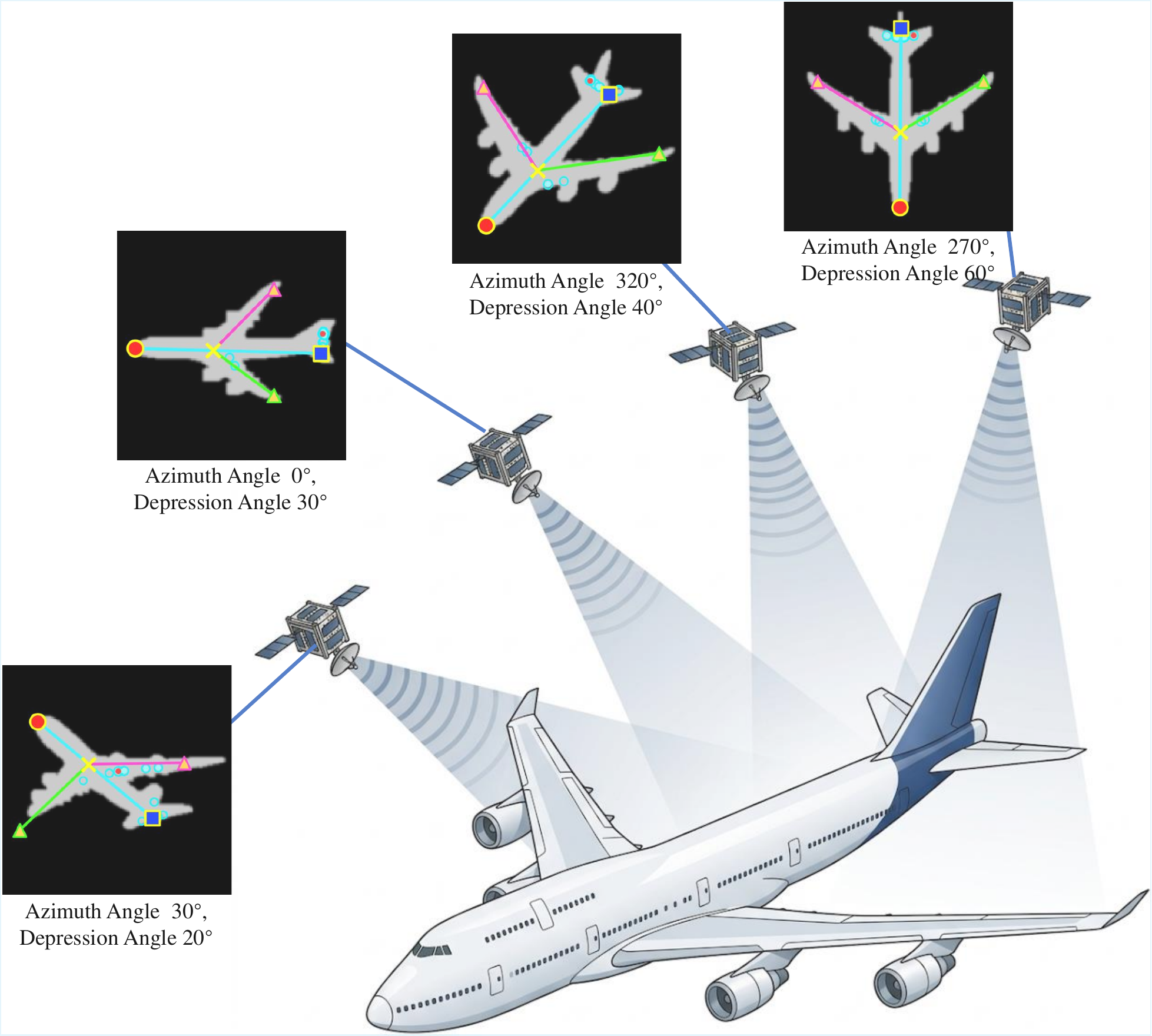}
\caption{Illustration of the 3D Pose Extraction results across varying radar viewpoints. As the azimuth and depression angles change, the projected GECM poses (e.g., nose, tail, and wingtips) remain consistent with the underlying 3D geometric transformation and radar imaging constraints, indicating that the extracted structural prior follows the corresponding viewpoint change.}
\label{fig:3d_pose_angles}
\end{figure}

Figure~\ref{fig:3d_pose_angles} illustrates the behavior of the 3D pose extraction module across varying radar viewpoints.
 
Simultaneously, to capture microscopic electromagnetic behavior, we perform geometric-optics multi-bounce ray tracing on $\mathcal{M}$. For each facet $f$ with centroid $\mathbf{c}_f$, normal $\mathbf{n}_f$, and area $A_f$, the visibility $v_f$ and local incidence angle $\theta_f$ are computed as:
\begin{equation}
\label{eq:visibility}
\begin{split}
v_f &= \max\!\big(0,\mathbf{n}_f^\top(-\hat{\mathbf{k}}_{\mathrm{i}})\big)\cdot \mathbf{1}_{\mathrm{vis}}(f),\\
\cos\theta_f &= \mathbf{n}_f^\top(-\hat{\mathbf{k}}_{\mathrm{i}}).
\end{split}
\end{equation}
For a traced path $\pi_m=(f_1,\ldots,f_m)$ up to $M$ bounces, we employ a practical energy-recursive scoring model (as opposed to coherent complex-path summation) for stable numerical implementation:
\begin{equation}
\label{eq:amplitude}
\begin{split}
E_t &= E_{t-1}\,R_t,\qquad \text{with}\quad E_0=1,\\
R_t &= \bar{\Gamma}_t\, \exp\!\left[-\left(\frac{4\pi \sigma_r \cos\theta_t}{\lambda}\right)^2\right] \max(\cos\theta_t, c_{\min}),
\end{split}
\end{equation}
where $\bar{\Gamma}_t$ is the Fresnel power response averaged over polarimetric components, $\sigma_r$ is the surface roughness scale, and $c_{\min}$ is a small cosine floor ensuring numerical stability. In our zero-shot setting, $\sigma_r$ and $\bar{\Gamma}_t$ are assigned empirically as homogeneous metallic constants across the CAD mesh. The path saliency is subsequently defined as:
\begin{equation}
\label{eq:saliency}
\omega(\pi_m)= \frac{E_m\,w_{\mathrm{align}}(\pi_m)\,\big(1+\mu(m-1)\big)} {\big(L_{\mathrm{in}}(\pi_m)+L_{\mathrm{out}}(\pi_m)\big)^2},
\end{equation}
where $w_{\mathrm{align}}(\pi_m)$ measures the backscatter-direction consistency, $L_{\mathrm{in}}$ and $L_{\mathrm{out}}$ are the inbound and outbound propagation distances respectively, and $\mu$ controls the multi-bounce gain.
 
\begin{figure*}[!t]
\centering
\includegraphics[width=\textwidth]{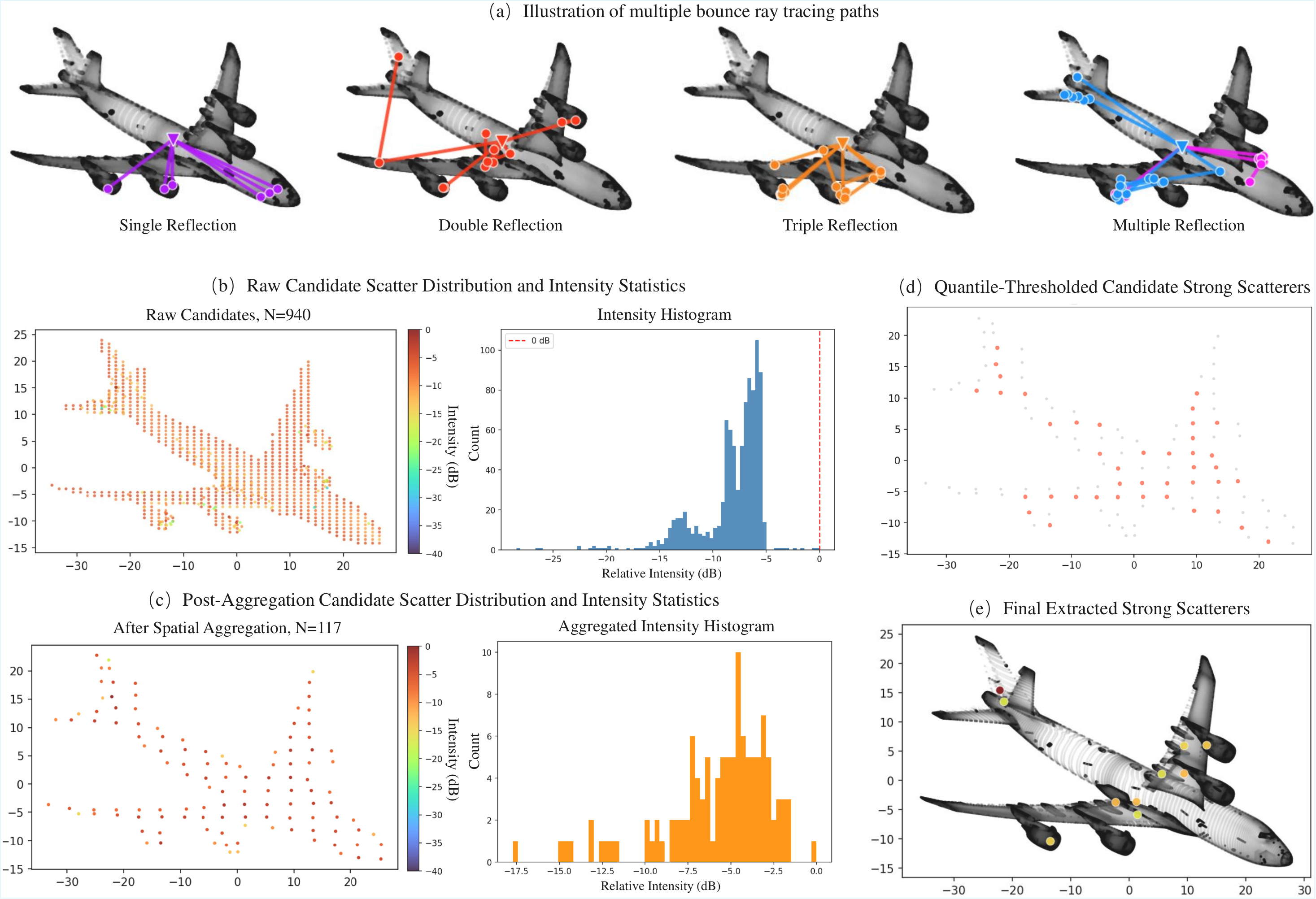}
\caption{Visualization of the intermediate processes within the Scattering Simulation branch. (a) Illustration of simulated multi-bounce ray tracing paths capturing complex electromagnetic interactions. (b) Distribution and intensity histogram of the raw candidate scatterers before processing ($N=940$). (c) Results after continuous image-plane spatial aggregation (grid binning), reducing dense ray-intersection jitter into regional saliency estimates ($N=117$). (d) Filtered candidate strong scatterers after relative decibel ($\mathrm{dB}$) thresholding. (e) The final extracted Top-$K$ representative physical scattering centers after NMS, overlaid on the 3D target.}
\label{fig:scattering_midresults}
\end{figure*}
 
To efficiently handle dense ray intersections and mitigate spurious artifacts, the raw bounce terminals are spatially aggregated via image-plane grid binning. Let $\Pi_{\boldsymbol{\eta}}:\mathbb{R}^3\!\to\!\mathbb{R}^2$ denote the continuous SAR image-plane projection under imaging parameters $\boldsymbol{\eta}$, and define $\mathbf{q}_i=(u_i,v_i)=\Pi_{\boldsymbol{\eta}}(\mathbf{x}_i)$ for the $i$-th ray terminal $\mathbf{x}_i\in\mathbb{R}^3$. Here, $(u_i,v_i)$ are \emph{continuous image-plane coordinates} (not mesh texture-UV coordinates). For explicit mathematical rigor, the accumulated saliency $\omega_i$ assigned to the terminal $\mathbf{x}_i$ is formalized as the sum of path saliencies from all traced paths ending at $\mathbf{x}_i$:
\begin{equation}
\label{eq:omega_i}
\omega_i=\sum_{\pi_m:\,\mathrm{term}(\pi_m)=\mathbf{x}_i}\omega(\pi_m).
\end{equation}
The resulting local candidates are then refined through relative logarithmic thresholding and non-maximum suppression (NMS):
\begin{equation}
\label{eq:candidates}
\begin{aligned}
\tilde{w}_k &= \sum_{\mathbf{q}_i \in \mathcal{B}_k} \omega_i, \\
\tilde{w}^{\mathrm{dB}}_k &= 10\log_{10}\!\left( \frac{\tilde{w}_k}{\max_j \tilde{w}_j} \right), \\
\tilde{\mathcal{P}}_{\mathrm{strong}}^{2d} &= \operatorname{TopK}\!\Big( \operatorname{NMS}\!\big(\big\{ \bar{\mathbf{q}}_k \mid \tilde{w}^{\mathrm{dB}}_k \ge \tau_{\mathrm{dB}} \big\}\big),\, K \Big),
\end{aligned}
\end{equation}
where $\mathcal{B}_k\subset\mathbb{R}^2$ represents the $k$-th continuous image-plane bin, and $\bar{\mathbf{q}}_k$ is its centroid. Here, $\tau_{\mathrm{dB}}$ serves as a relative decibel threshold to filter weak clutter returns, and the output is explicitly capped to the top-$K$ candidates.
 
The spatial aggregation stage in Eq.~(\ref{eq:candidates}) is a dedicated energy consolidation operation before sparsification. By accumulating multiple ray terminals within each local bin $\mathcal{B}_k$, it transforms point-wise, sampling-density-sensitive responses into a more stable regional saliency estimate $\tilde{w}_k$. This reduces sensitivity to ray-intersection jitter and mitigates the bias caused by redundant local oversampling around the same physical scattering center. Importantly, spatial aggregation and NMS are complementary rather than redundant: the former performs local energy integration, while the latter enforces non-redundant representative selection among surviving candidates.
 
To illustrate the microscopic scattering simulation pipeline, Fig.~\ref{fig:scattering_midresults} visualizes the step-by-step intermediate results of the electromagnetic branch. Starting from the dense and noisy raw multi-bounce ray tracing terminals [Fig.~\ref{fig:scattering_midresults}(a) and Fig.~\ref{fig:scattering_midresults}(b)], the spatial aggregation mechanism consolidates the energy and mitigates intersection jitter [Fig.~\ref{fig:scattering_midresults}(c)]. Subsequent decibel thresholding and NMS further refine these responses [Fig.~\ref{fig:scattering_midresults}(d)], ultimately isolating representative physical scattering centers [Fig.~\ref{fig:scattering_midresults}(e)] for discrete rasterization.

To bridge physical continuous coordinates with the discrete generation canvas, we further introduce a rasterization operator $\mathcal{R}_{\gamma}:\mathbb{R}^2\!\to\!\mathbb{Z}^2$ determined by spatial resolution $\gamma$. The effective projection used by GECM rendering can be written as:
\begin{equation}
\label{eq:sar_projection_composition}
\mathcal{P}_{\mathrm{sar}}(\cdot;\boldsymbol{\eta})=\mathcal{R}_{\gamma}\!\circ\!\Pi_{\boldsymbol{\eta}}(\cdot).
\end{equation}
For branch-wise clarity, we explicitly define the pixel-domain outputs as:
\begin{subequations}
\label{eq:branch_pixel_sets}
\begin{align}
\mathbf{P}_{\text{pose}}^{pix}
&=\mathcal{R}_{\gamma}\!\left(\Pi_{\boldsymbol{\eta}}\!\left(\mathbf{P}_{\text{pose}}^{3d}\right)\right),\\
\mathbf{P}_{\text{scat}}^{pix}
&=\mathcal{R}_{\gamma}\!\left(\tilde{\mathcal{P}}_{\text{strong}}^{2d}\right).
\end{align}
\end{subequations}
Eq.~(\ref{eq:branch_pixel_sets}) makes explicit that the pose branch uses projection followed by rasterization, while the strong-scatterer branch is rasterized directly from continuous image-plane coordinates. This eliminates reliance on real target masks during inference and preserves true zero-shot capability.
 
For clarity, the final semantic GECM tensor $\mathbf{G}\in\mathbb{R}^{H\times W\times 3}$ is formed by the channel-aware composition of the rasterized pose and strong-scatterer primitives, utilizing the identical symbolic encoding established in the training phase (cyan fuselage, green/pink wings, red/blue markers, and intensity-coded scatterers):
\begin{equation}
\label{eq:gecm_comp}
\mathbf{G}
=
\mathcal{E}_{\mathrm{pose}}\!\left(\mathbf{P}_{\text{pose}}^{pix}\right)
\oplus
\mathcal{E}_{\mathrm{strong}}\!\left(\mathbf{P}_{\text{scat}}^{pix},\,\tilde{w}\right).
\end{equation}
Specifically, the rendering intensity for strong points is logarithmically scaled:
\begin{equation}
\label{eq:intensity}
I(\mathbf{x}_{2d})=\frac{\log\!\big(1+\kappa\,\tilde{w}(\mathbf{x}_{2d})\big)}{\log\!\big(1+\kappa\,\tilde{w}_{\max}\big)}\in[0,1].
\end{equation}
 
\begin{figure*}[!t]
\centering
\includegraphics[width=\linewidth]{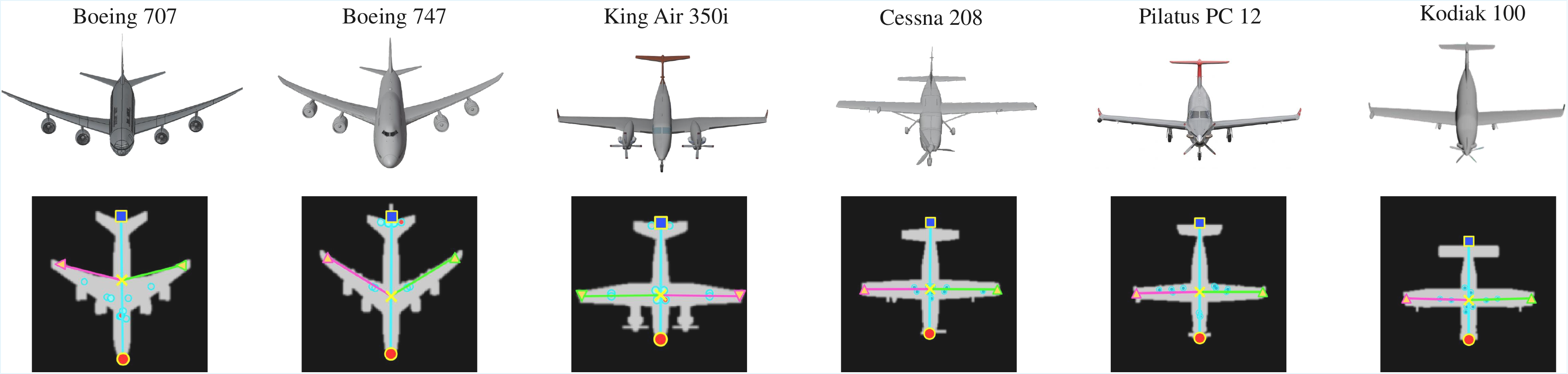}
\caption{Zero-shot generalization of the 3D-guided GECM generation across diverse aircraft CAD models (e.g., Boeing 707, Boeing 747, King Air 350i, Cessna 208, Pilatus PC 12, Kodiak 100), showing that the forward generation process does not rely on real SAR masks.}
\label{fig:3dgecm_models}
\end{figure*}

The fundamental necessity of this 3D guidance stems from the inverse ambiguity inherent in SAR imaging. For an observed SAR image $\mathbf{Y}$ and its underlying 3D geometry $\mathbf{X}$:
\begin{equation}
\label{eq:forward_model}
\mathbf{Y}=\mathcal{F}(\mathbf{X};\boldsymbol{\eta})+\mathbf{n}.
\end{equation}
Accurately recovering $\mathbf{X}$ from $\mathbf{Y}$ alone is heavily underconstrained. Our method addresses this limitation by introducing a deterministic, forward-simulated structural prior:
\begin{equation}
\label{eq:prior}
\mathbf{G}=g(\mathcal{M},\boldsymbol{\eta}),
\end{equation}
which subsequently conditions the diffusion generation process:
\begin{equation}
\label{eq:diffusion}
\hat{\mathbf{v}}_t=\mathcal{F}_{\theta,\psi,\phi}(\mathbf{z}_t,t,\mathbf{T},\mathbf{G}).
\end{equation}
Consequently, the proposed 3D-guided GECM generation process serves as a geometric-electromagnetic bridge that constrains the generated SAR image to remain consistent with both the target viewpoint and the underlying scattering mechanism. Its applicability is not limited to a single aircraft category: as shown in Fig.~\ref{fig:3dgecm_models}, the same forward generation process can produce consistent GECMs for multiple aircraft CAD models under zero-shot settings.

\subsection{Decoupled Training and Zero-Shot Inference Strategy}
\label{subsec:training_inference_strategy}
 
Based on the unified GECM representation, we adopt a decoupled conditional generation strategy: the training phase learns the mapping from semantic control to realistic SAR appearance using real SAR data, while the inference phase replaces the real-image-derived control with a 3D-model-guided GECM generated from a CAD model and specified imaging parameters. Because both phases share the same GECM space and rendering semantics, the learned prior can transfer naturally to unseen targets.
 
\paragraph{Training phase}
For each real SAR sample, we construct a paired triplet
\begin{equation}
\left(I_i,\mathbf{T}_i,\mathbf{G}_i^{\mathrm{tr}}\right),
\end{equation}
where $I_i$ is the real SAR image, $\mathbf{T}_i=\tau(\boldsymbol{\eta}_i)$ is the text condition derived from SAR metadata, and $\mathbf{G}_i^{\mathrm{tr}}=h(I_i,\boldsymbol{\eta}_i)$ is the corresponding GECM derived from the real image. The image is encoded into latent space as
\begin{equation}
\mathbf{z}_{0,i}=\mathcal{E}_{\mathrm{ae}}(I_i),
\end{equation}
and corrupted by Gaussian noise:
\begin{equation}
\mathbf{z}_{t,i}=(1-\sigma_t)\mathbf{z}_{0,i}+\sigma_t\boldsymbol{\epsilon},
\qquad
\boldsymbol{\epsilon}\sim\mathcal{N}(\mathbf{0},\mathbf{I}).
\end{equation}
 
The conditional denoising model is written as
\begin{equation}
\hat{\mathbf{v}}_{t,i}
=
\mathcal{F}_{\theta,\psi,\phi}
\left(
\mathbf{z}_{t,i},\, t,\, \mathbf{T}_i,\, \mathbf{G}_i^{\mathrm{tr}}
\right),
\end{equation}
where $\theta$ denotes the FLUX backbone, $\psi$ the LoRA parameters, and $\phi$ the ControlNet parameters. LoRA adapts the semantic feature space associated with SAR parameters and text conditions, while ControlNet injects the GECM to enforce pose structure and scattering layout. Following the flow-matching formulation, the regression target is
\begin{equation}
\mathbf{v}_{t,i}^{\star}=\boldsymbol{\epsilon}-\mathbf{z}_{0,i},
\end{equation}
and the training objective is
\begin{equation}
\mathcal{L}_{\mathrm{train}}
=
\mathbb{E}
\left[
\left\|
\mathcal{F}_{\theta,\psi,\phi}
\left(
\mathbf{z}_{t,i}, t, \mathbf{T}_i, \mathbf{G}_i^{\mathrm{tr}}
\right)
-
\mathbf{v}_{t,i}^{\star}
\right\|_2^2
\right].
\end{equation}
 
\paragraph{Inference phase}
At inference time, no real SAR image is required. Given a CAD model $\mathcal{M}$ and specified SAR imaging parameters $\boldsymbol{\eta}$, we first generate a zero-shot GECM
\begin{equation}
\mathbf{G}^{\mathrm{zs}}=g(\mathcal{M},\boldsymbol{\eta}),
\end{equation}
and construct the corresponding text condition
\begin{equation}
\mathbf{T}^{\mathrm{zs}}=\tau(\boldsymbol{\eta}).
\end{equation}
The GECM controls viewpoint-dependent structure and strong-scatter distribution, while the text condition specifies global SAR attributes such as class, band, polarization, and resolution.
 
Starting from Gaussian latent noise $\mathbf{z}^{(0)}\sim\mathcal{N}(\mathbf{0},\mathbf{I})$, the trained FLUX+LoRA+ControlNet model performs iterative denoising:
\begin{equation}
\mathbf{z}^{(k+1)}
=
\mathbf{z}^{(k)}
+
\Delta t_k\,
\mathcal{F}_{\theta,\psi,\phi}
\left(
 \mathbf{z}^{(k)},\, t_k,\, \mathbf{T}^{\mathrm{zs}},\, \mathbf{G}^{\mathrm{zs}}
\right),
\end{equation}
followed by latent decoding
\begin{equation}
\hat{I}=\mathcal{D}_{\mathrm{ae}}\!\left(\mathbf{z}^{(K)}\right).
\end{equation}
This process is zero-shot because the target need not appear in the real SAR training set. Real SAR data provide the appearance prior during training, while LoRA and ControlNet preserve semantic and geometric control at inference.

\section{Experiments}
\label{sec:experiments}

\subsection{Experimental Setup}
\label{subsec:experimental_setup}

\subsubsection{Dataset Description}
\begin{figure*}[!t]
\centering
\includegraphics[width=\textwidth]{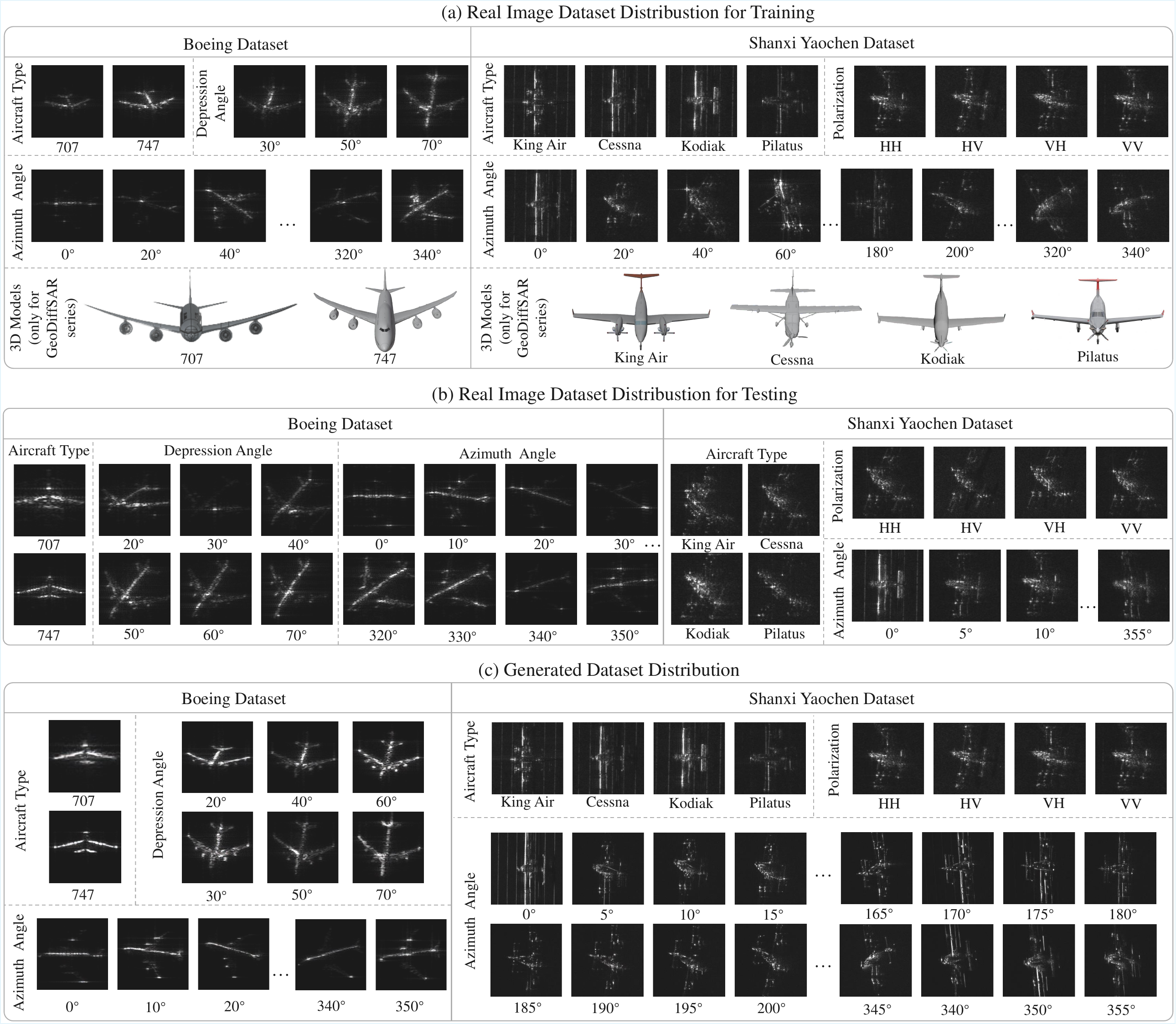}
\caption{Detailed illustration of the dataset partitioning and content for both the simulated Boeing dataset and the real-world Shanxi Yaocheng dataset. The visualization shows the sparsity of the training sets (limited azimuths, depression angles, and polarizations) relative to the denser testing sets. The training data is utilized for generative model optimization and baseline downstream ATR training, while the complete testing sets are exclusively reserved for metric evaluation and zero-shot downstream classification testing.}
\label{fig:dataset_intro}
\end{figure*}

We evaluate GeoDiff-SAR II on two SAR datasets under a unified sparse-view protocol: a simulated Boeing dataset and a real Shanxi dataset. For each dataset, all methods are trained on the same real training split and evaluated on the same held-out real test split. After training, each method synthesizes samples under the test conditions. Image-quality metrics are computed against the held-out real test split. For downstream ATR evaluation, the generated samples are merged with the real training split to train a classifier, which is tested exclusively on the held-out real test split. Figure~\ref{fig:dataset_intro} summarizes the partitioning.

\paragraph{Boeing Dataset}
The Boeing dataset contains high-fidelity simulated SAR images of Boeing 707 and Boeing 747 aircraft generated by the Suzhou Research Institute, Aerospace Information Research Institute, Chinese Academy of Sciences (AIR CAS). The sparse training split contains 74 images sampled every $20^\circ$ in azimuth and only three depression angles ($30^\circ$, $50^\circ$, and $70^\circ$), leaving $20^\circ$, $40^\circ$, and $60^\circ$ entirely unseen during training. The held-out test split densifies the sampling to azimuth intervals of $10^\circ$ and depression-angle intervals of $10^\circ$ from $20^\circ$ to $70^\circ$. Accordingly, each method must synthesize the missing viewpoints on the full $36\times 6$ azimuth-depression-angle grid.

\begin{figure*}[!t]
\centering
\includegraphics[width=\textwidth]{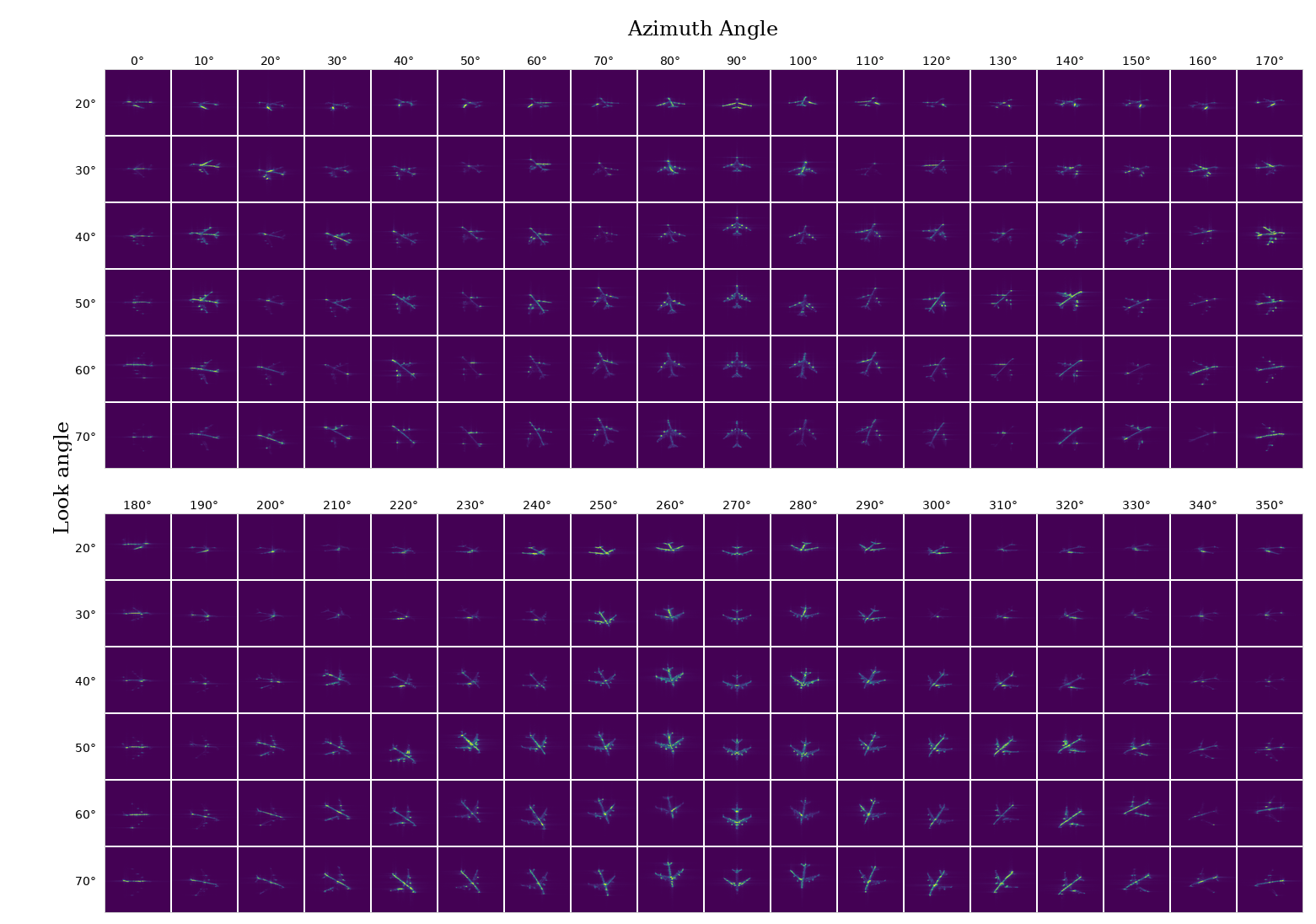}
\caption{Comprehensive zero-shot generation results of the Boeing 747 aircraft using the GeoDiff-SAR II framework. The grid displays synthesized SAR images spanning full continuous azimuth angles ($0^\circ$--$350^\circ$) and full depression angles ($20^\circ$--$70^\circ$). The results at unseen depression angles (e.g., $20^\circ, 40^\circ, 60^\circ$) show that the model can generalize across missing viewpoints while remaining consistent with the imposed 3D geometric constraints.}
\label{fig:allangle}
\end{figure*}

Full-grid zero-shot synthesis results on the Boeing dataset are shown in Fig.~\ref{fig:allangle}.

\paragraph{Shanxi Dataset}
The Shanxi dataset contains real SAR images collected at Yaocheng Airport, Shanxi, China, covering four aircraft classes: Cessna 208, King Air 350i, Kodiak 100, and Pilatus PC 12. The sparse training split includes 1,546 images sampled every $20^\circ$ in azimuth across all four polarization modes (HH, HV, VH, VV). The held-out test split contains 6,666 images sampled every $5^\circ$, so intermediate azimuths absent from training dominate the evaluation set. The augmentation target covers all 72 azimuth bins under the four polarization modes.

\subsection{Visual Quality Evaluation}
\label{subsec:visual_quality_evaluation}

We evaluate synthesized SAR images using PSNR, SSIM, LPIPS, and FID. Higher PSNR and SSIM and lower LPIPS and FID indicate better agreement with the held-out real test data. For each method, samples are generated under the test conditions. PSNR, SSIM, and LPIPS are computed against semantically matched held-out real samples, while FID compares the generated set with the held-out real test distribution. Tables~\ref{tab:boeing_metrics} and \ref{tab:shanxi_metrics} report the quantitative results.
The compared baselines include VQGAN-CLIP \cite{crowson2022vqganclip}, DeepFloyd IF \cite{deepfloyd2023if}, PixArt-$\alpha$ \cite{chen2023pixart}, SDXL \cite{podell2023sdxl}, DALL-E \cite{ramesh2022hierarchical}, SD3.5-Medium \cite{stability2024sd35}, FLUX 1.0 \cite{blackforest2024flux}, and GeoDiff-SAR (V1) \cite{zhang2026geodiff}.

\begin{table}[t]
\centering
\caption{Quantitative image quality metrics on the Boeing dataset. All methods are evaluated on the held-out test split.}
\label{tab:boeing_metrics}
\scriptsize
\begin{tabular*}{\linewidth}{@{\extracolsep{\fill}} l c c c c @{}}
\toprule
\textbf{Method} & \textbf{PSNR $\uparrow$} & \textbf{SSIM $\uparrow$} & \textbf{LPIPS $\downarrow$} & \textbf{FID $\downarrow$} \\
\midrule
VQGAN-CLIP~\cite{crowson2022vqganclip} & 29.70 & 0.22 & 0.23 & 274.54 \\
DeepFloyd IF~\cite{deepfloyd2023if} & 27.07 & 0.32 & 0.19 & 116.00 \\
PixArt-$\alpha$~\cite{chen2023pixart} & 31.36 & 0.55 & 0.18 & 75.71 \\
SDXL~\cite{podell2023sdxl} & 30.67 & 0.23 & 0.14 & 165.39 \\
DALL-E~\cite{ramesh2022hierarchical} & 21.45 & 0.21 & 0.24 & 111.93 \\
SD3.5-Medium~\cite{stability2024sd35} & 25.32 & 0.64 & 0.16 & 59.89 \\
FLUX 1.0~\cite{blackforest2024flux} & \underline{32.68} & \underline{0.89} & \underline{0.06} & 49.48 \\
GeoDiff-SAR (V1)~\cite{zhang2026geodiff} & 31.40 & 0.83 & 0.07 & \underline{30.20} \\
\textbf{GeoDiff-SAR II} & \textbf{35.26} & \textbf{0.97} & \textbf{0.02} & \textbf{16.35}\\
\bottomrule
\end{tabular*}
\end{table}

\begin{table}[t]
\centering
\caption{Quantitative image quality metrics on the Shanxi dataset. All methods are evaluated on the held-out test split.}
\label{tab:shanxi_metrics}
\scriptsize
\begin{tabular*}{\linewidth}{@{\extracolsep{\fill}} l c c c c @{}}
\toprule
\textbf{Method} & \textbf{PSNR $\uparrow$} & \textbf{SSIM $\uparrow$} & \textbf{LPIPS $\downarrow$} & \textbf{FID $\downarrow$} \\
\midrule
VQGAN-CLIP~\cite{crowson2022vqganclip} & \underline{23.87} & 0.19 & 0.39 & 242.22 \\
DeepFloyd IF~\cite{deepfloyd2023if} & 20.13 & 0.21 & 0.58 & 191.57 \\
PixArt-$\alpha$~\cite{chen2023pixart} & 21.74 & 0.44 & 0.38 & 80.80 \\
SDXL~\cite{podell2023sdxl} & 21.01 & 0.16 & 0.49 & 192.06 \\
DALL-E~\cite{ramesh2022hierarchical} & 21.86 & 0.63 & 0.38 & 91.86 \\
SD3.5-Medium~\cite{stability2024sd35} & 23.40 & 0.67 & 0.33 & 78.47 \\
FLUX 1.0~\cite{blackforest2024flux} & 23.54 & 0.64 & 0.52 & 61.77 \\
GeoDiff-SAR (V1)~\cite{zhang2026geodiff} & 23.68 & \underline{0.76} & \underline{0.31} & \underline{36.27} \\
\textbf{GeoDiff-SAR II} & \textbf{29.26} & \textbf{0.86} & \textbf{0.25} & \textbf{18.02} \\
\bottomrule
\end{tabular*}
\end{table}

Among the compared methods, GeoDiff-SAR II attains the lowest FID and LPIPS on Boeing and the highest PSNR together with the lowest FID on Shanxi. These results indicate that explicit GECM conditioning improves both local fidelity and global distributional alignment. Compared with FLUX 1.0 \cite{blackforest2024flux} and GeoDiff-SAR (V1) \cite{zhang2026geodiff}, the results also suggest that backbone capacity alone is insufficient under sparse-view SAR generation, whereas explicit geometric-electromagnetic conditioning provides more reliable performance, particularly on the cluttered Shanxi dataset.

\begin{figure*}[!t]
\centering
\includegraphics[width=\textwidth]{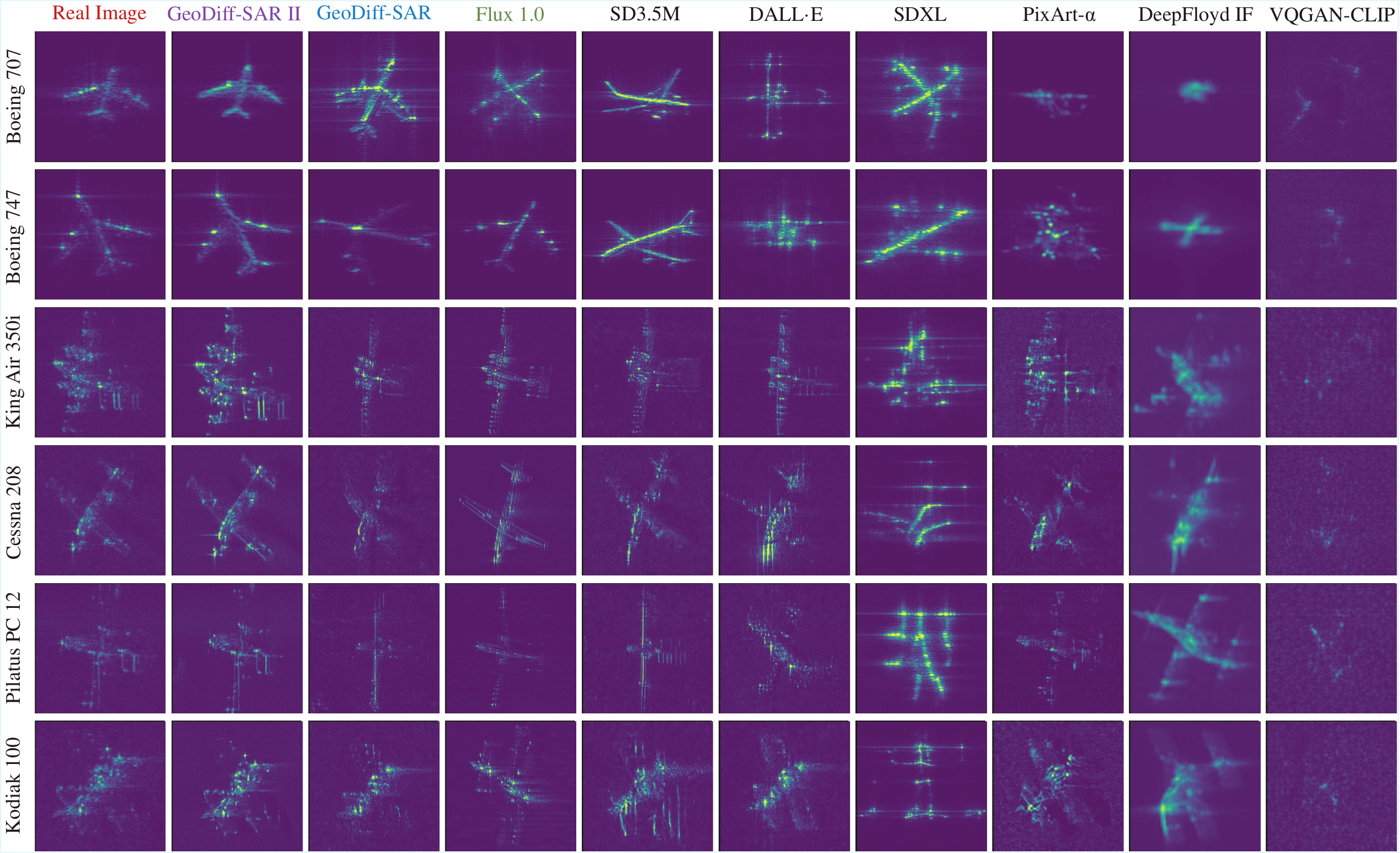}
\caption{Qualitative visual comparison of generated SAR images across different methods under challenging zero-shot conditions. From top to bottom, the specific parameters for each row are: (1) Boeing 707 (Depression: $40^\circ$, Azimuth: $100^\circ$); (2) Boeing 747 (Depression: $50^\circ$, Azimuth: $70^\circ$); (3) King Air 350i (Polarization: HH, Azimuth: $15^\circ$); (4) Cessna 208 (Polarization: HV, Azimuth: $305^\circ$); (5) Pilatus PC 12 (Polarization: VV, Azimuth: $5^\circ$); (6) Kodiak 100 (Polarization: VH, Azimuth: $130^\circ$). Testing views such as the $40^\circ$ depression angle and azimuths of $5^\circ, 15^\circ$ are absent from the training sets. Compared with the baselines, GeoDiff-SAR II exhibits fewer physically inconsistent artifacts and closer agreement with the real reference images.}
\label{fig:ex-cmpfig}
\end{figure*}

\paragraph{Qualitative Visual Comparison}
Representative qualitative comparisons are shown in Fig.~\ref{fig:ex-cmpfig}.

\subsection{Experiments on Downstream Classification Tasks}
\label{subsec:downstream_classification}

To evaluate the practical utility of the synthesized images, we train a standard downstream CNN classifier on the real training split augmented by the generated samples from each method. The classifier is then evaluated exclusively on the held-out real test split. Macro-averaged Precision, Recall, and F1-scores are reported in Table~\ref{tab:boeing_downstream_macro} (Boeing) and Table~\ref{tab:shanxi_downstream_macro} (Shanxi).

\paragraph{Boeing Dataset}
Table~\ref{tab:boeing_downstream_macro} shows that training only on the sparse real split yields 23.01\% depression-angle F1 and 1.16\% azimuth F1, reflecting the limited viewpoint coverage in the baseline supervision. Generic image-generation baselines provide limited gains, with azimuth F1 remaining below 10\% for most methods. In contrast, GeoDiff-SAR II raises the depression-angle and azimuth F1-scores to \textbf{97.46\%} and \textbf{90.05\%}, compared with 18.08\% and 8.55\% for FLUX 1.0 \cite{blackforest2024flux} and 14.58\% and 7.10\% for GeoDiff-SAR (V1) \cite{zhang2026geodiff}. This indicates that the generated samples contribute viewpoint continuity consistent with the underlying SAR geometry, rather than merely increasing sample quantity.

\begin{table*}[!t]
\centering
\caption{Downstream classification results on the simulated Boeing dataset (macro precision/recall/F1, \%).}
\label{tab:boeing_downstream_macro}
\resizebox{\linewidth}{!}{
\begin{tabular}{ l c c c c c c c c c }
\toprule
\multirow{2}{*}{\textbf{Method}} & \multicolumn{3}{c}{\textbf{Target (Aircraft)}} & \multicolumn{3}{c}{\textbf{Depression Angle}} & \multicolumn{3}{c}{\textbf{Azimuth Angle}} \\
\cmidrule(lr){2-4} \cmidrule(lr){5-7} \cmidrule(lr){8-10}
& \textbf{Precision} & \textbf{Recall} & \textbf{F1-Score} & \textbf{Precision} & \textbf{Recall} & \textbf{F1-Score} & \textbf{Precision} & \textbf{Recall} & \textbf{F1-Score} \\
\midrule
True Data (Baseline)& 80.00 & 76.90 & 76.29 & 16.79 & \underline{40.49} & \underline{23.01} & 1.14  & 1.96  & 1.16 \\
VQGAN-CLIP~\cite{crowson2022vqganclip} & 84.69 & 80.70 & 80.13 & 14.24 & 35.57 & 19.89 & 4.84  & 11.69 & 6.32 \\
DeepFloyd IF~\cite{deepfloyd2023if} & 79.90 & 66.37 & 62.09 & 8.67  & 23.59 & 12.50 & 3.91  & 12.70 & 5.22 \\
PixArt-$\alpha$~\cite{chen2023pixart} & 84.48 & 77.49 & 76.28 & 13.44 & 34.88 & 18.72 & 4.73  & 10.50 & 6.05 \\
SDXL~\cite{podell2023sdxl} & 81.22 & 73.68 & 71.99 & 5.91  & 24.69 & 9.53  & 0.34  & 3.97  & 0.61 \\
DALL-E~\cite{ramesh2022hierarchical} & 79.38 & 64.91 & 59.99 & 14.80 & 25.70 & 15.37 & 4.64  & 14.68 & 6.02 \\
SD3.5-Medium~\cite{stability2024sd35} & 72.83 & 71.93 & 71.65 & 10.96 & 24.06 & 13.28 & 0.75  & 4.35  & 1.22 \\
FLUX 1.0~\cite{blackforest2024flux} & 83.75 & 83.33 & 83.28 & 15.98 & 31.47 & 18.08 & \underline{13.89} & \underline{15.05} & \underline{8.55} \\
GeoDiff-SAR (V1)~\cite{zhang2026geodiff} & \underline{90.12} & \underline{88.89} & \underline{88.80} & \underline{17.52} & 25.63 & 14.58 & 6.64  & 12.29 & 7.10 \\
\textbf{GeoDiff-SAR II} & \textbf{94.53} & \textbf{93.86} & \textbf{93.84} & \textbf{97.47} & \textbf{97.48} & \textbf{97.46} & \textbf{91.30} & \textbf{90.00} & \textbf{90.05} \\
\bottomrule
\end{tabular}
}
\end{table*}

Table~\ref{tab:boeing_downstream_macro} also shows that the gain in viewpoint recognition does not come at the expense of target semantics. Methods with visually plausible textures still do not yield comparable gains on depression-angle and azimuth classification when their generated samples do not preserve physically correct viewpoint evolution. This helps explain why generic generators remain below the proposed method on the two physically meaningful attributes. More broadly, the Boeing results suggest that SAR augmentation should provide a continuous and physically consistent viewpoint manifold rather than isolated visual realism. GeoDiff-SAR II improves classification mainly in regions where the real training set is sparse or missing, indicating that its generated samples carry transferable geometric structure.

\begin{figure*}[!t]
\centering
\includegraphics[width=\textwidth]{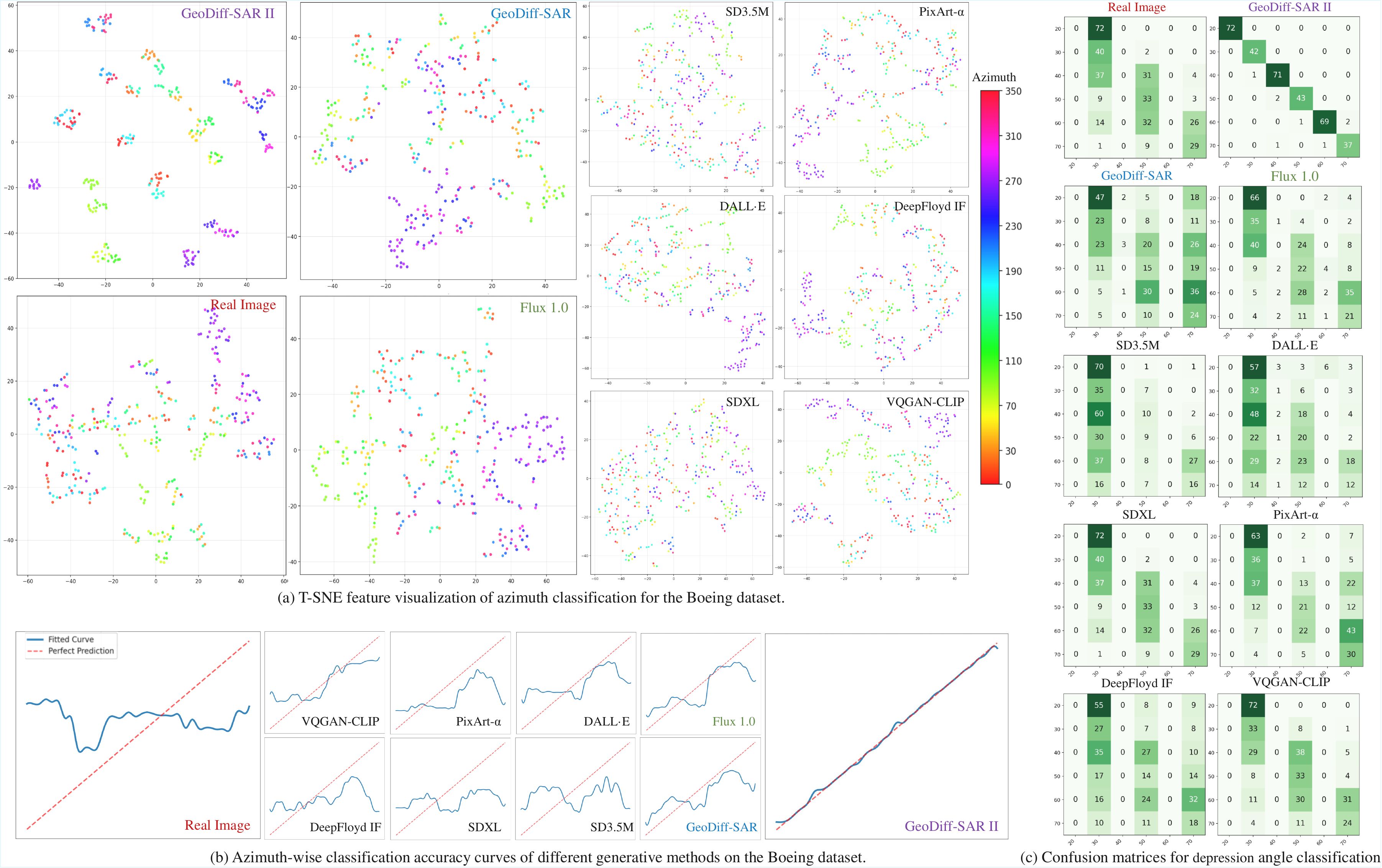}
\caption{Visualizations of downstream ATR performance on the simulated Boeing dataset. (a) t-SNE visualization of learned feature distributions for azimuth classification. (b) Azimuth-wise classification accuracy curves across varying angles. (c) Confusion matrices for depression-angle classification. Compared with the baselines, GeoDiff-SAR II reduces the distribution gaps at unseen depression angles (e.g., $20^\circ, 40^\circ, 60^\circ$) and yields improved classification performance.}
\label{fig:ex_classify_boeing}
\end{figure*}

\paragraph{Shanxi Dataset}
Table~\ref{tab:shanxi_downstream_macro} reflects a more pronounced distribution shift. The true-data baseline achieves only 2.46\% azimuth F1, and generic foundation-model baselines remain below the proposed method on both polarization and azimuth recognition. GeoDiff-SAR II improves the polarization and azimuth F1-scores to \textbf{94.32\%} and \textbf{95.01\%}, showing that explicit 3D-guided GECM conditioning transfers effectively to real sparse-view SAR data. This result is obtained even though the Shanxi test split is dominated by intermediate azimuths absent from training.

\begin{table*}[!t]
\centering
\caption{Downstream classification results on the real-world Shanxi dataset (macro precision/recall/F1, \%).}
\label{tab:shanxi_downstream_macro}
\resizebox{\linewidth}{!}{
\begin{tabular}{ l c c c c c c c c c }
\toprule
\multirow{2}{*}{\textbf{Method}} & \multicolumn{3}{c}{\textbf{Target (Aircraft)}} & \multicolumn{3}{c}{\textbf{Polarization Mode}} & \multicolumn{3}{c}{\textbf{Azimuth Angle}} \\
\cmidrule(lr){2-4} \cmidrule(lr){5-7} \cmidrule(lr){8-10}
& \textbf{Precision} & \textbf{Recall} & \textbf{F1-Score} & \textbf{Precision} & \textbf{Recall} & \textbf{F1-Score} & \textbf{Precision} & \textbf{Recall} & \textbf{F1-Score} \\
\midrule
True Data (Baseline)& 93.75 & 93.03 & 93.36 & 74.54 & 75.34 & 74.59 & 1.31  & 23.57 & 2.46 \\
VQGAN-CLIP~\cite{crowson2022vqganclip} & 92.16 & 88.98 & 90.03 & 75.05 & 74.60 & 74.81 & 1.39  & 23.30 & 2.59 \\
DeepFloyd IF~\cite{deepfloyd2023if} & 94.47 & 91.14 & 92.49 & 73.39 & 72.94 & 73.00 & 1.59  & 22.60 & 2.82 \\
PixArt-$\alpha$~\cite{chen2023pixart} & 98.14 & 96.67 & 97.32 & 76.93 & 76.80 & 74.98 & \underline{40.19} & \underline{37.80} & \underline{20.68} \\
SDXL~\cite{podell2023sdxl} & 88.85 & 79.54 & 79.77 & 64.74 & 63.87 & 62.73 & 1.26  & 22.72 & 2.37 \\
DALL-E~\cite{ramesh2022hierarchical} & 98.91 & 98.84 & 98.87 & \underline{80.65} & \underline{80.73} & \underline{80.63} & 12.00 & 25.25 & 5.67 \\
SD3.5-Medium~\cite{stability2024sd35} & \underline{99.53} & \underline{99.43} & \underline{99.48} & 79.76 & 79.99 & 79.85 & 31.63 & 33.31 & 16.98 \\
FLUX 1.0~\cite{blackforest2024flux} & 98.39 & 97.91 & 98.15 & 77.61 & 77.93 & 77.65 & 2.26  & 23.13 & 2.84 \\
GeoDiff-SAR (V1)~\cite{zhang2026geodiff} & 99.15 & 99.28 & 99.21 & 79.88 & 80.10 & 79.78 & 20.09 & 25.28 & 6.40 \\
\textbf{GeoDiff-SAR II} & \textbf{99.80} & \textbf{99.71} & \textbf{99.76} & \textbf{94.34} & \textbf{94.32} & \textbf{94.32} & \textbf{95.27} & \textbf{96.33} & \textbf{95.01} \\
\bottomrule
\end{tabular}
}
\end{table*}

\begin{figure*}[!t]
\centering
\includegraphics[width=\textwidth]{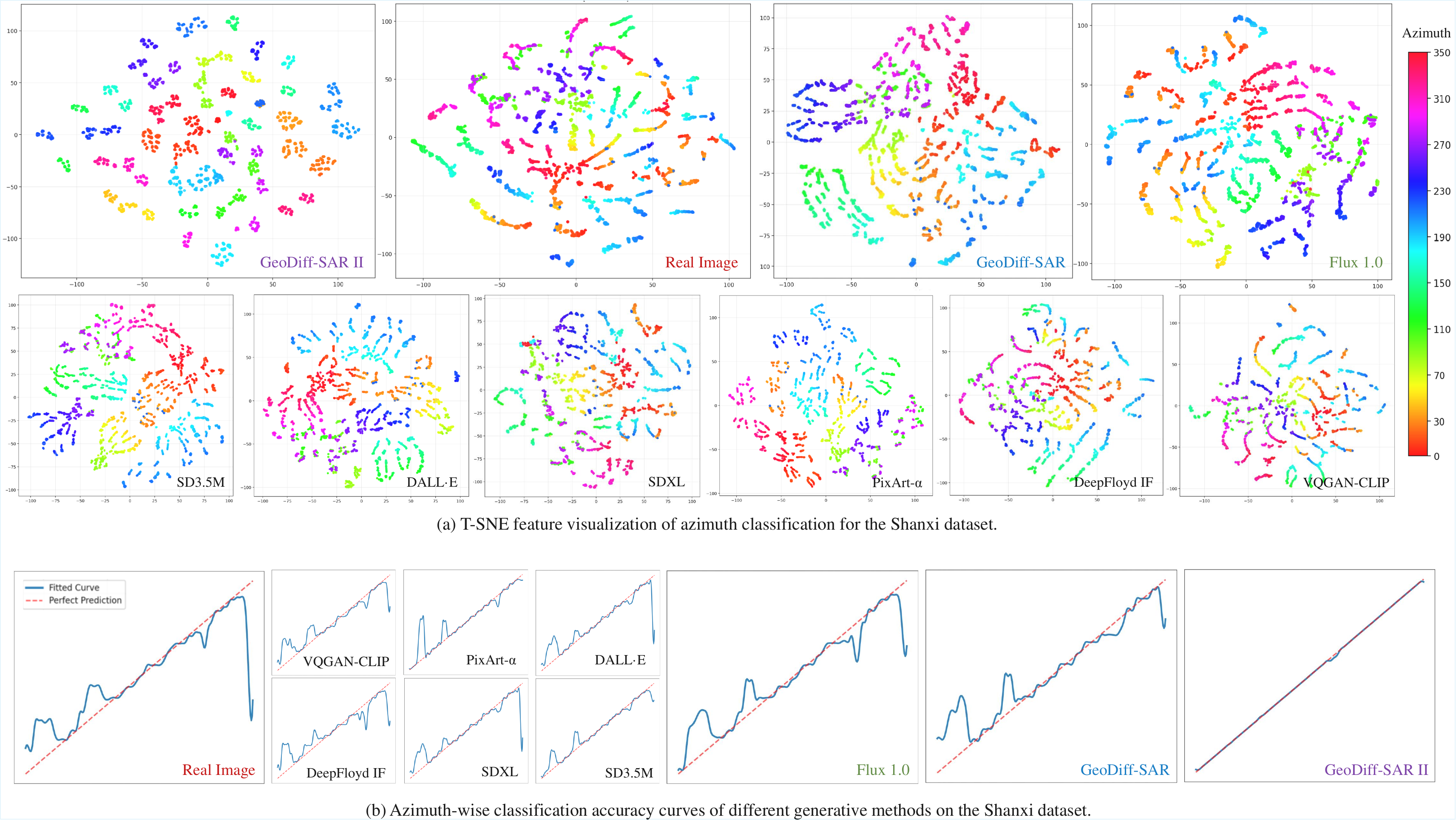}
\caption{Visualizations of downstream ATR performance on the real-world Shanxi dataset. (a) t-SNE feature visualization for azimuth classification, showing clearer feature separation under the proposed method. (b) Azimuth-wise classification accuracy curves. GeoDiff-SAR II follows the reference prediction curve more closely, indicating improved robustness to the distribution shift between the sparse training azimuths and the dense testing azimuths.}
\label{fig:ex_classify_shanxi}
\end{figure*}

Table~\ref{tab:shanxi_downstream_macro} further shows that the proposed method continues to provide gains when augmentation is learned from real, noisy SAR observations rather than purely simulated data. Compared with the Boeing setting, the Shanxi scene introduces stronger clutter, denser speckle, and a larger mismatch between sparse training azimuths and dense testing azimuths. The performance margin of GeoDiff-SAR II indicates that the proposed conditioning strategy retains pose and polarization cues relevant to SAR imaging despite the more complex appearance distribution. Although some generic generators recover partial azimuth trends or moderate target-level discrimination, their gains remain fragmented across attributes. In contrast, the proposed method shows gains across all three evaluation dimensions, suggesting closer alignment with the underlying SAR formation process. Figure~\ref{fig:ex_classify_shanxi} supports this interpretation: the t-SNE manifolds show clearer separation across azimuths, and the azimuth-wise accuracy curve remains close to the reference trajectory over the full angle range.

\section{Conclusion}
This paper presents GeoDiff-SAR II, a 3D model-guided decoupled framework for controllable SAR image generation. By introducing the GECM, the proposed method unifies SAR-image-derived geometric-electromagnetic supervision during training with 3D-model-driven physical rendering during inference, thereby enabling explicit control over azimuth angle, depression angle, and polarization mode within a single generation pipeline. Experiments on both the simulated Boeing dataset and the real Shanxi dataset demonstrate that the proposed method maintains stable performance across large azimuth gaps and yields consistent gains in image fidelity, physical consistency, and downstream ATR performance over representative baselines. These findings indicate that explicit geometric-electromagnetic conditioning provides a practical approach to SAR data augmentation when multi-view real data are limited. Future work will extend the framework toward more complex targets, broader sensing configurations, and tighter integration between physical simulation and generative foundation models.

\section*{Acknowledgment}
This work was supported in part by the National Natural Science Foundation of China under Grant Nos. 62271034 and 62331026, the Natural Science Foundation of Shandong Province under Grant ZR2024ZD19, and the Fundamental Research Funds for the Central Universities under Grant Nos. ZY2610, ZY2609, and ZY2614.

\end{document}